\documentclass[10pt]{iopart}
\usepackage{graphicx}
\usepackage{subfig}
\usepackage{xcolor}
\usepackage{amssymb}
\usepackage{setstack}
\usepackage{arydshln}
\usepackage[square,sort&compress, comma, numbers]{natbib}
\usepackage{enumitem}
\usepackage{multirow}
\usepackage{url}

\definecolor{ctecolor}{rgb}{0.05, 0.2, 0.8}
\definecolor{lnkcolor}{rgb}{0,0,0.6} 
\definecolor{refcolor}{rgb}{0.95,0.3,0}
\definecolor{cyan}{rgb}{0.302, 0.745, 0.933}
\definecolor{cccyan}{rgb}{0.271, 0.890, 1}
\definecolor{ooorange}{rgb}{0.851, 0.325, 0.098}
\definecolor{pppurple}{rgb}{0.243, 0.149, 0.659}
\definecolor{yyyellow}{rgb}{0.976, 0.984, 0.082}
\definecolor{gggreen}{rgb}{0.459, 0.980, 0.141}
\definecolor{rrred}{rgb}{0.980, 0.569, 0.388}
\definecolor{cccyan2}{rgb}{0, 0.212, 0.612}
\definecolor{gggreen2}{rgb}{0.059, 0.380, 0}
\definecolor{rrred2}{rgb}{0.651, 0.031, 0.149}

\begin{document}

\title{Emissive cathode immersed in a plasma: plasma-cathode interactions, operation and stability}
\author{F. Pagaud$^1$, V. Dolique$^1$, N. Claire$^2$ and N. Plihon$^1$}
\address{$^1$Univ Lyon, ENS de Lyon, CNRS, Laboratoire de Physique, F-69342 Lyon, France}
\address{$^2$Aix Marseille Univ, CNRS, PIIM, Marseille, France}
\begin{abstract}
Thermionic emission from a polycrystalline tungsten emissive cathode immersed in a magnetized plasma column is investigated experimentally and numerically. Electrical and optical measurements of the cathode temperature show a highly inhomogeneous cathode temperature profile due to plasma-cathode interactions. The spatially and temporally resolved cathode temperature profile provides an in-depth understanding of the thermionic electron current, in excellent agreement with experimental data. The plasma-cathode coupling leads to a sharp and heterogeneous rise in temperature along the cathode, which can eventually lead to unstable cathode operation, with divergent current growth. A detailed thermal modeling accurately reproduces the experimental measurements, and allows to quantify precisely the relative importance of heating and cooling mechanisms in the operation of the cathode immersed in the plasma. Numerical resolution of the resulting integro-differential equation highlights the essential role of heterogeneous ohmic heating and the importance of ion bombardment heating in the emergence of unstable regimes. Detailed thermal modelling enables operating regimes to be predicted in excellent agreement with experimental results.
\end{abstract}

\maketitle
\ioptwocol

\section{Introduction}

Electron emission from thermionic \textit{hot-cathodes} has been routinely used as a source of primary electrons to ionize a plasma since the pioneering works of Edison, as reported by Preece~\cite{Preece_1885}, and of Fleming on vacuum tubes~\cite{Fleming_1904}. Thermionic emission refers to electron emission from negatively biased surfaces heated above typically $2000~$K.  Richardson demonstrated that the thermionic current follows an Arrhenius-like law~\cite{Richardson_1901}, subsequently corrected and now called Richardson's law (see Eq.\ref{eq:i_eth}). In 1928, the importance of thermionic emission in technological applications was underlined by the award of the Nobel Prize award to Richardson \textit{"for his work on the thermionic phenomenon and especially for the discovery of the law named after him"}.

The use of hot cathodes is not limited to vacuum tubes, but was and still is widely used for primary electron production in plasma sources. Several regimes of operation have been reported for glow discharges in the presence of hot cathodes~\cite{Bosch_1986,Greiner_1993,arnas_capeau_analysis_1996,Pae_2002}. Transitions, oscillations and multistability between these different regimes are understood from the complex interactions between the plasma and the cathode. The modeling of the interaction between the plasma and emissive cathodes is still an active research topic~\cite{Campanell_2017,pedrini_theoretical_2015,Moritz_2023}. Emissive cathodes were also widely used as primary sources for 
Q machines~\cite{Saeki_1979}, moderate size~\cite{matsukuma_spatiotemporal_2003,Brochard_2005} and very large~\cite{gekelman_design_1991, gekelman_upgraded_2016} linear magnetized plasma columns, toroidal machines~\cite{Rypdal_1994,Prasad_1994,Greiner_2003,Alex_2022,Lemoine_2005}, smaller dedicated  experiments~\cite{makrinich_2009}, or as a source for neutralizing and ionising electrons in space-propulsion systems~\cite{goebel_lab6_2007,vincent_electron_2020}. 
In the context of plasma transport control in tokamaks, biasing experiments using emissive cathodes demonstrated improved confinement and the emergence of a transport barrier by current injection, more than 40 years ago~\cite{Taylor_1989}. Edge biasing using emissive surfaces is still an active line of research in the fusion community, leading to  improved confinement~\cite{Silva_2004} or the  control of runaway electrons~\cite{Ghanbari_2012}. The behavior of thermionic emissive plasma-facing components has several potential advantages~\cite{Tolias_2020}. A first potential application is the operation of emissive divertor in the presence of an inverse sheath (an electron-rich sheath), that favors plasma detachment and low target plasma temperatures~\cite{Campanell_2020}. 
A second advantage of thermionic emission lies in the cooling of components in contact with the plasma, due to both radiation and cooling by the emitted electrons, which carry away the potential barrier energy associated to the emission~\cite{Komm_2017}. Cooling from thermionic emission has also been proposed for the atmospheric re-entry of spacecrafts, known in this context as electron transpiration cooling~\cite{Uribarri_2015,Hanquist_2017}.

There has been a recent surge of interest in the study of plasma flow induced by the interaction of large current emitted by hot cathodes with magnetic fields. For instance, the Big Red Ball~\cite{Forest_2015} and the Plasma Couette eXperiment~\cite{Flanagan_2020} at the University of Wisconsin implement the interaction of large current injection from emissive cathodes with large-scale or multipolar magnetic fields to study laboratory astrophysical relevant phenomena such as the dynamo instability, the magnetorotational instability or the dynamic of  the Parker spiral in the solar wind~\cite{Peterson_2019}. Hot emissive cathodes immersed in a pre-existing plasma also shed new light on transport and plasma turbulence in magnetized plasma columns~\cite{DuBois_2014,Gilmore_2015}, plasma flow generation~\cite{desangles_bousselin_poye_plihon_2021, jin_plasma_2019}, or the dynamic of interacting plasma filaments~\cite{sydora_drift-alfven_2019, karbashewski_magnetized_2022}.
The control of electric fields perpendicular to the ambient magnetic field is also crucial for a number of applications of $E\times B$ configurations~\cite{Kaganovich_2020}, and among them, high-throughput plasma mass separation~\cite{gueroult_opportunities_2018,Zweben_2018}. The control of the plasma potential and of the plasma rotation using current injection from emissive cathode have recently been the subject of experimental investigations~\cite{pagaud_sub2023,Liziakin_2021} and of theoretical modeling~\cite{ trotabas_trade-off_2022, poulos_model_2019}, which require a precise description of the cathode behavior in the presence of a plasma. 

\begin{figure*}[tbp]
    \begin{center}
    \includegraphics[width = 0.95\textwidth]{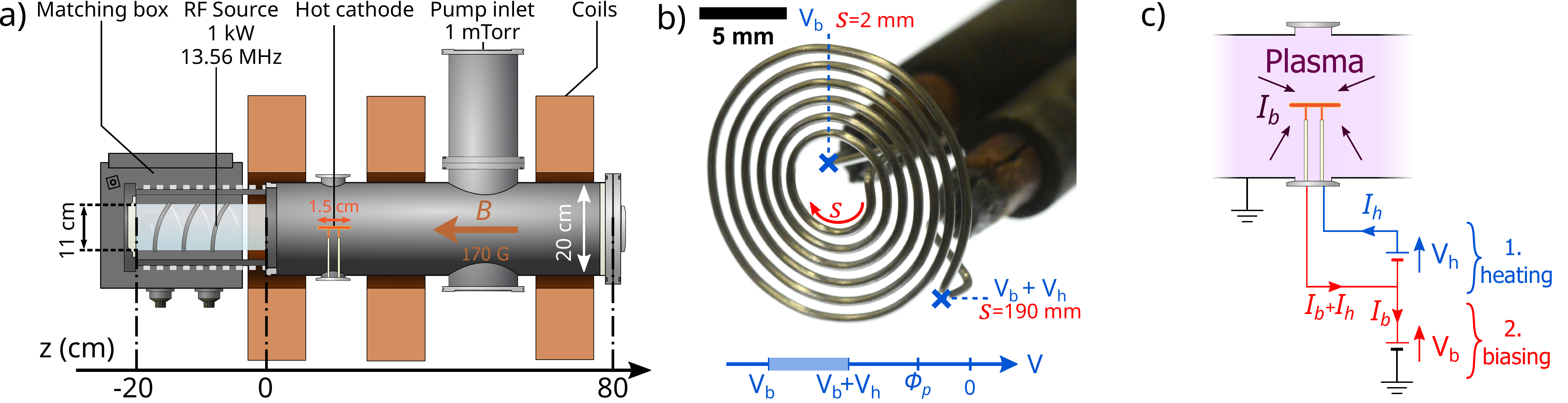}
    \caption{\textit{a)} Sketch of the experimental setup. \textit{b)} Photograph of the tungsten emissive cathode. \textit{c)} Electrical circuit of the electrode. 
     }
    \label{fig:vkp}
    \end{center}
\end{figure*}
This brief review shows that there are still a number of cutting-edge technological applications involving thermionic cathode immersed in a pre-existing plasma. Most of the understanding of their functioning relies on established physical properties (see for the instance the review~\cite{herring_thermionic_1949} or the recent review on oxides and borides emissive materials~\cite{gao_review_2020, taran_review_2009}) and extensive thermophysical properties dataset~\cite{kawano_effective_2022}. Nevertheless, the operation of a hot cathode immersed in a pre-existing plasma requires a careful modeling of the plasma-cathode interaction. In this article, we report an experimental study and a detailed modeling of the operation of a hot tungsten cathode immersed in a high-density magnetized plasma column. Spatially and temporally resolved temperature measurements allow to precisely predict the cathode current, and a thermal budget is developed. 
This article is organized as follows. In Sec.~\ref{sec:setup} the experimental setup is briefly introduced and measurements of the cathode temperature in presence of plasma are presented in Sec.~\ref{sec:experimental}. The dynamics of a unsteady regime is then characterized experimentally in Sec.~\ref{sec:empirical}, using spatially and temporally resolved measurements of the cathode temperature. An energy budget equation presented in Sec.~\ref{sec:simulation}, taking into account the plasma-cathode interactions, is in excellent agreement with experimental characterization of the dynamics of the unsteady regime. This model allows to explain the observations and predict the regimes of operations of the cathode in Sec. \ref{sec:operation}.

\section{Plasma source and cathode description}
\label{sec:setup}

\subsection{The von-K\'arm\'an plasma experiment}

The Von-K\'{a}rm\'{a}n Plasma (VKP) experiment consists of a magnetized plasma column sketched in \fref{fig:vkp}(a) and described in details
in~\citep{plihon_flow_2015}. The plasma is created  by a $13.56~\mathrm{MHz}$ inductive antenna made of a 3-turns helicoidal coil wrapped around a 11 cm wide borosilicate tube and fed through a L-type matching network. Argon gas is injected through a puffed valve at the top of the experiment at $z=16$ cm, close to the source tube, and is pumped down by a primary turbo-molecular pump located at $z = 49$ cm. For the set of experiments presented in this article, the plasma pressure is 1 mTorr and the radio-frequency forward power is $1~\mathrm{kW}$.  The plasma then expands in a 80 cm long, 20 cm in diameter grounded stainless-steel cylinder, and is confined by an axial magnetic field created by a
set of three Bitter coils, of amplitude $B = 170~\mathrm{G}$. The end disks of the cylindrical vessel are insulating (borosilicate or boron-nitride disks). Plasma columns generated in the VKP experiment have typical radii of 5 cm, plasma densities in the $10^{18} \mathrm{m}^{-3}$ range and electron temperature around 4 eV~\cite{desangles_bousselin_poye_plihon_2021,vincent_high-speed_2022}. 
Due to thermal constraints, plasma shots of a few seconds are pulsed with a typical repetition rate of 60 seconds. The level of shot to shot reproducibility
is estimated to be 0.6\% for the ion saturation current of a Langmuir probe, with a standard deviation of 0.2\% (estimated from a series of 40 shots).

\subsection{Operation of an additional  emissive cathode}

A hot emissive cathode is inserted in the plasma column from a lateral port, at $z=16$ cm. It consists in a 6-turns spiral-shaped filament of pure polycristalline tungsten  of radius $r_W = 253~\mathrm{\mu m}$, and total length $l_W = 198~\mathrm{mm}$~\cite{desangles_bousselin_poye_plihon_2021} (see \fref{fig:vkp}(b))). The outer diameter of the spiral is 1.5 cm. The curvilinear coordinate $s$ is defined with $s=2$ mm at the center of the spiral (note the mounting base of the filaments that are not in the spiral plane). 
The cathode is operated using two independent electrical DC power supplies (see \fref{fig:vkp}(c)):

\begin{itemize}
    \item[1.] The tungsten filament is Joule heated by a given \textit{heating} current $I_h$, with no constraints on the voltage drop $V_h$;
    \item[2.] Strong thermionic emission is achieved setting a negative bias $V_b$, below the plasma potential.
\end{itemize}
\noindent
The total cathode current $I_b$ is limited  to $15~\mathrm{A}$ (note that, for the sake of simplicity, the convention is such that $I_b > 0$ when electrons are emitted from the cathode). 

\begin{figure}[tp]
    \centering
    \includegraphics[width= 0.48\textwidth]{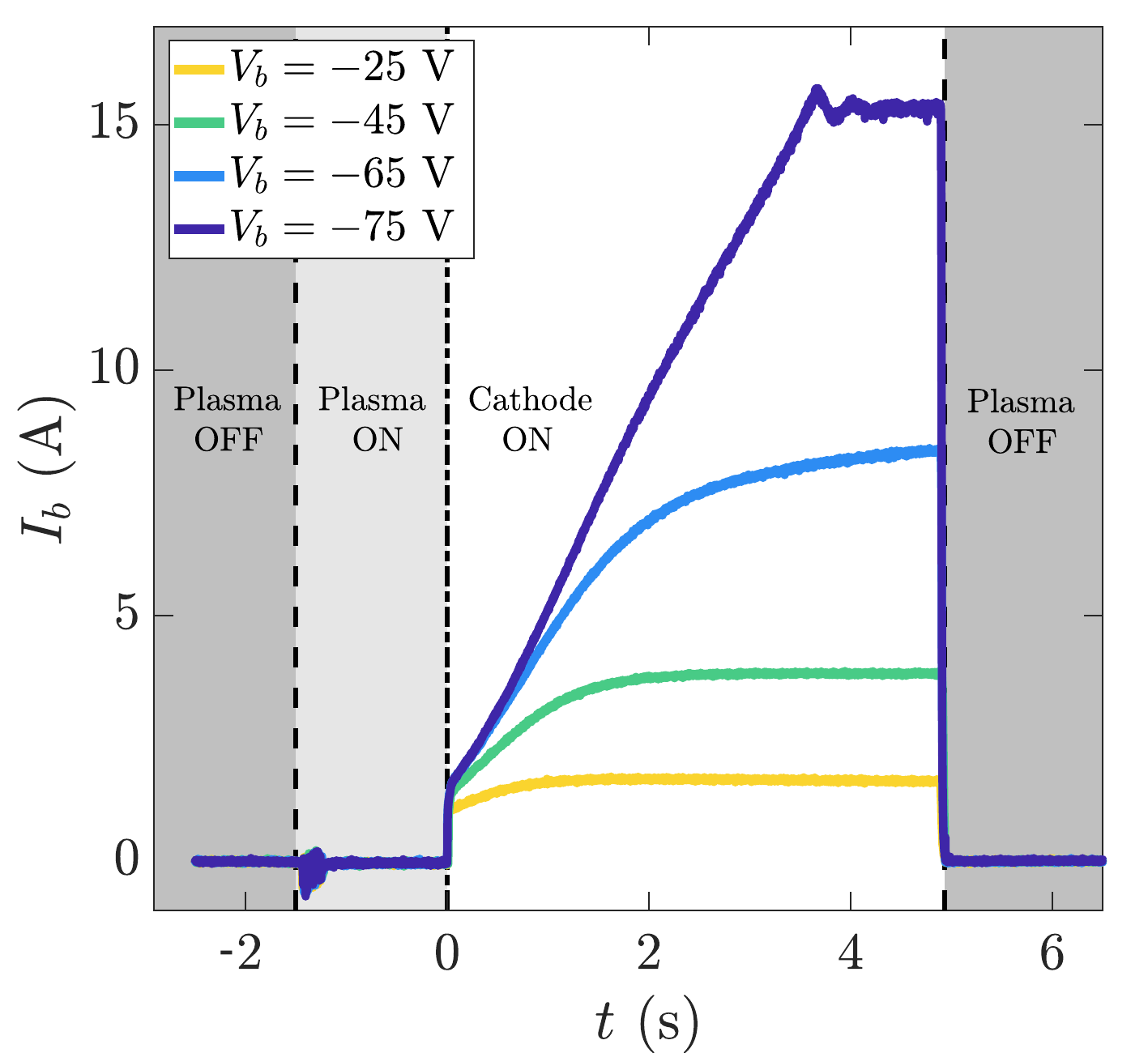}
    \caption{Time evolution of the emitted current $I_b$ for a given heating current $I_h$ and various values of the cathode bias $V_b$.}
    \label{fig:div_regime}
\end{figure}

Typical time series of the  current $I_b$ during plasma shots are shown in Fig. \ref{fig:div_regime}. The cathode is continuously heated, whether the radio-frequency plasma is ignited or not, and  the cathode bias is left floating initially. The plasma is ignited at $t=-1~$s and the cathode is negatively biased at $t=0~$s. A strong cathode current $I_b$ is then emitted starting at time $t=0~$s, until the plasma is turned off ($t=5~$s for the shots shown in Fig.~\ref{fig:div_regime}). Note that the bias voltage  $V_b$ is referenced to ground and feds the center of the spiral (see Fig.~\ref{fig:div_regime}) and that the heating current $I_h$ is set constant whether the plasma is ignited or not (here $I_h = 17~\mathrm{A}$). A naive interpretation of the above-described protocol is the following: the cathode temperature is set by the heating power supply, independently of the cathode bias. Since, for a given cathode material, thermionic emission is controlled by the cathode temperature, this should allow independent control of the injected current and potential. However, as observed in Fig.~\ref{fig:div_regime}, the more negative the voltage bias, the higher the injected current. As shown later, this current increase is linked to an inhomogeneous increase of the cathode temperature due to  plasma-cathode interactions, as the bias becomes increasingly negative. The goal of this article is to provide precise estimates of the cathode current from detailed measurements of the cathode temperature profile and its modeling. Steady-state regimes are reached for low bias voltages ($V_b = ~–25~\mathrm{V}$ and $–45~\mathrm{V}$ in Fig.~\ref{fig:div_regime}) and \textit{divergent} regimes, with runaway evolution of $I_b$ are observed at larger bias voltage ($V_b = ~–75~\mathrm{V}$ in Fig.~\ref{fig:div_regime}). Note that the current $I_b$ was hardware-limited to $15~A$ to avoid any damage to the cathode, but the model developed in Sec.~\ref{sec:simulation} shows that, when $V_b = ~–75~\mathrm{V}$, the cathode experiences an unbounded temperature growth leading to current divergence.

Let us now recall the features of electron emission for an emissive cathode. The upper bound of thermionic emission \cite{herring_thermionic_1949} is set by the Richardson's law:

\begin{equation}
    I_{em} = A_\mathrm{g}\mathcal{A}T_W^2\exp\left(-\frac{eW}{k_\mathrm{B}T_W}\right)
\label{eq:i_eth}
\end{equation}
\noindent
with $A_\mathrm{g} = 6\times 10^5 ~\mathrm{A/(K^2~m^2)}$ the Richardson constant for tungsten, $\mathcal{A} = 314~\mathrm{mm^2}$ the surface of the tungsten filament, $e$ the elementary charge, $W = 4.54~\mathrm{eV}$ the work function of tungsten and $k_\mathrm{B}$ the Boltzmann constant. 
We consider here that the emitted current reaches the Richardson current, which is observed in the absence of space-charge limitations~\cite{ye_effect_2000, cavalier_strongly_2017, poulos_model_2019}, \textit{i.e.} when the pre-existing plasma density is high enough and the cathode potential $V_b$ is well below the plasma potential $\Phi_p$. 
The current  $I_b$ delivered by the bias power supply, is the sum of the emitted current $I_{em}$, the ion saturation current $I_{is}$ and the electron current $I_e$ collected by the cathode at potential $V_b$: 
\begin{eqnarray}
I_b = I_{em} + I_{is}~–~I_e \label{eq:i_b}\\
I_{is} = \mathcal{A}ne\sqrt{\frac{eT_e}{m_i}} \label{eq:i_is}\\
I_e = I_{is}\exp\left(\Lambda + \frac{V_b~–~\Phi_p}{T_e}\right) \label{eq:i_e}
\end{eqnarray}
\noindent with $m_i$ and $m_e$ being respectively the ion and electron masses and $\Lambda = \ln\left(\sqrt{2m_i/(\pi m_e)}\right)$ a sheath parameter. The plasma parameters (plasma density $n$ and electron temperature $T_e$) facing the cathode are assumed to be homogeneous. The potential and kinetic contributions of ions to secondary-electron emission should be low before $I_{is}$~\cite{konuma_film_1992, tolias_secondary_2014} and are therefore neglected.

The amplitude of the Richardson current depends on the cathode temperature $T_W$ and the work function $W$. We assume $W$ to be independent of $T_W$, despite a lack of consensus in the literature~\cite{durakiewicz_thermal_2001, kawano_effective_2008} (the evolution of $W$ is bounded by $15~\mathrm{meV}$ between $2500~K$ and $3000~K$). The electric field at the cathode surface is estimated around $10^4~\mathrm{V/m}$, leading to variations of $W$ bounded by a few meV because of the Schottky effect~\cite{herring_thermionic_1949}, which can be ignored.

\section{Experimental measurements of the cathode temperature}
\label{sec:experimental}

Since the amplitude of the emitted current is set by the value of the temperature profile $T_W(s)$ of the cathode, precise temperature  estimates are essential. Two methods are investigated and compared in this section. 

\subsection{Effective temperature estimate from global electrical resistance measurement}
\label{sec:effective}

Temperature estimates from measurement of filaments resistance is a widely spread method, which relies on the temperature evolution of the resistivity of the materials. The temperature evolution of pure Tungsten  resistivity $\rho_W$ (corrected for thermal expansion) is accurately known~\cite{desai_electrical_1984, white_thermophysical_1997}, with a best power-law fit in the range [1800$~\mathrm{K}$; 3200$~\mathrm{K}$]:
\begin{equation}
    \rho_W = 5.31\times 10^{–11} T_W^{1.222}~–~1.56\times 10^{–9}~\mathrm{\Omega/m}
    \label{eq:rho}
\end{equation}
A direct measurement of the cathode resistance $R = V_h / I_h$, in the absence of emitted current, thus provides an estimate of an effective cathode temperature $\overline{T_W}$ as: 
\begin{equation}
    \overline{T_W} = \left[\left(\frac{\pi r_W^2}{l_W}R + 1.56\times 10^{–9}\right)\frac{10^{11}}{5.31}\right]^{1/1.222}.
    \label{eq:T}
\end{equation}
This method is particularly suited for cases where the temperature is uniform along the filament. In the case of current emission from the cathode, this estimate is only possible before the shot or at the end of the shot, and is not suitable to provide time-resolved measurements. Note that a contact resistance $R_\mathrm{con}$ between the filaments and the holder has to be considered; in our case, an empirical contact resistance of $10~\mathrm{m\Omega}$ was measured (typically 1\% of $R$). Moreover, we will show that in the case of strongly inhomogeneous temperature profiles, the estimate of the 
 current using $\overline{T_W}$ is strongly inaccurate.

\subsection{Optical method}

\begin{figure}[tbp]
    \begin{center}
    \includegraphics[width = 0.48\textwidth]{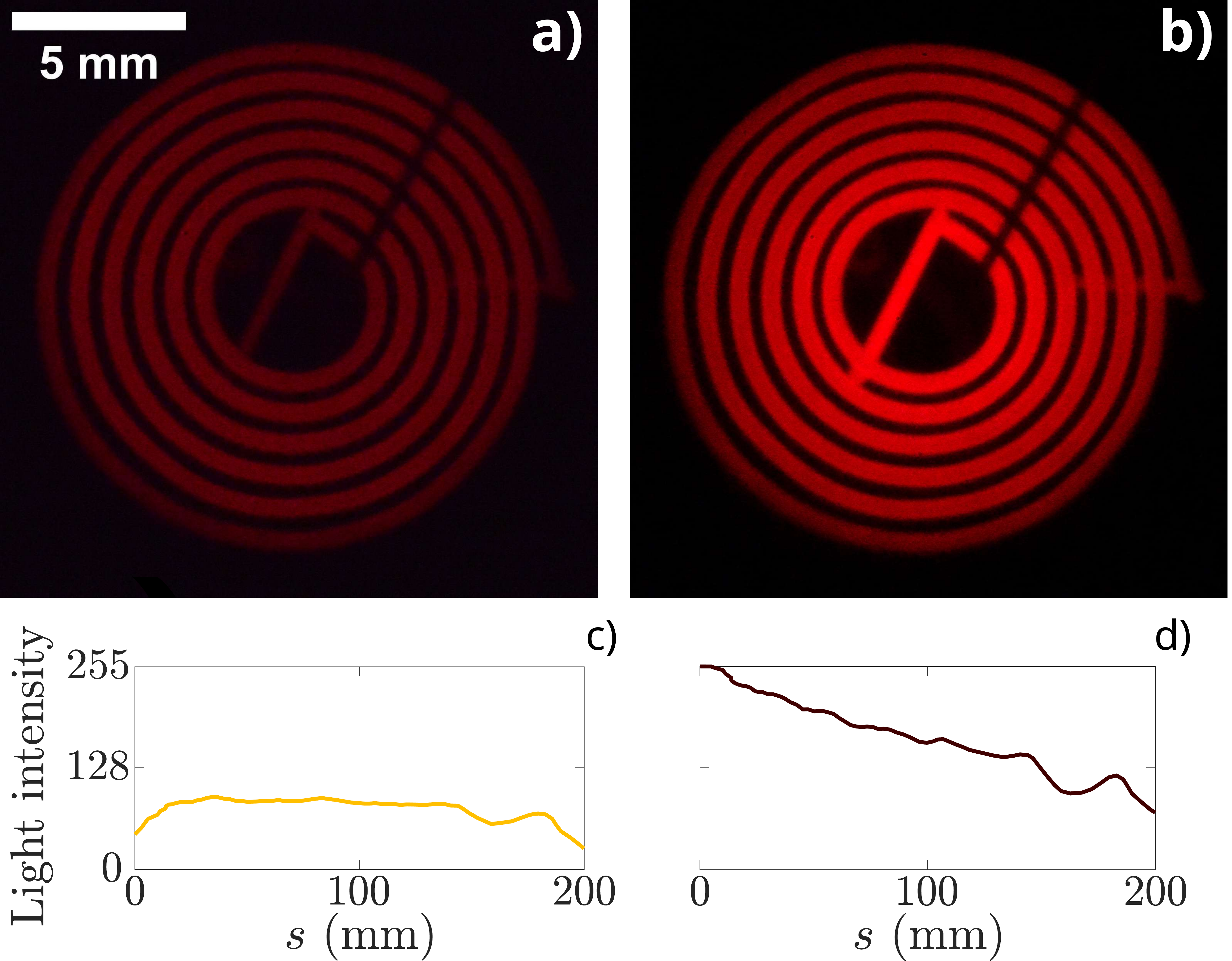}
    \caption{Hot cathode ($I_h = 16.3~\mathrm{A}$) observed through the pyrometer at 650 nm \textit{a)} without plasma and \textit{b)} in presence of plasma ($V_b = –61~\mathrm{V}$, $I_b = 8~\mathrm{A}$). The reference gray body is the darker line in the right upper corner. \textit{c)-d)} Corresponding light intensity profiles.}
    \label{fig:photo_cath_plasma}
    \end{center}
\end{figure}

Measurement of the filament temperature using an intensity comparison pyrometer is a powerful alternative method. The principle of operation relies on the comparison of the intensity of infrared radiation at a given wavelength, between a gray-body under test and a precisely calibrated reference filament. 
We used a Keller ITS Mikro PV 11 pyrometer, based on the measurement of the radiation at $650~$nm. The pyrometer is located 2.30 m away from the cathode and images the cathode through a borosilicate window. Images of the pyrometer output were shot using a Nikon D610 camera with a minimal resolution of 1920x1080 pixels, at a framerate of 30 fps. Typical snapshots of the cathode intensity at $650~$nm are shown in Fig.~\ref{fig:photo_cath_plasma}, before plasma ignition (\textit{i.e.} $t<-1$~s) and at the end of a plasma shot. The bottom panels of Fig.~\ref{fig:photo_cath_plasma} shows the evolution of the light intensity, which depends upon the local temperature $T_W$, along the curvilinear coordinate of the spiral. The  cathode temperature profile is clearly inhomogeneous after being biased in the plasma.

The quantitative interpretation of the radiation intensity relies on Stefan-Boltzmann law, corrected by a material-dependent emissivity factor $\epsilon$ that depends on wavelength $\lambda$ and $T_W$:
\begin{equation}
    P_{\sigma, \lambda} = \sigma \mathcal{A}\epsilon(\lambda, T_W) T_W^4, 
    \label{eq:stefan}
\end{equation}
with $\sigma = 5.7\times 10^{–8}~\mathrm{W/(m^2~K^4)}$ the Stefan-Boltzmann constant. 
The principle of the pyrometer exploits Eq.~\eref{eq:stefan} by matching the cathode light intensity at $650~$nm to the one from a reference gray body of tunable emissivity $\epsilon(\lambda = 650~\mathrm{nm}, T_W)$. Polycristalline tungsten emissivity at $650~$nm decreases with temperature $T_W$, from  $\epsilon_W(\lambda = 650~\mathrm{nm}, T_W =2200~\mathrm{K} ) = 0.44$ to $\epsilon_W(\lambda = 650~\mathrm{nm}, T_W =2800~\mathrm{K} ) = 0.425$~\cite{cezairliyan_simultaneous_1996, de_vos_new_1954}. As the cathode is imaged through a borosilicate window, whose transmittance $\tau_\mathrm{g}$ was measured to be 
$0.80\pm0.04$ at $650~$nm (see Supp. Mat. A), a global emissivity $\epsilon = \epsilon_W \tau_\mathrm{g}$ has to be considered. We chose a linear evolution of $\epsilon$ from 0.35 at $2000~$K to 0.325 at $3000~$K. A conversion factor from pixel intensity to local temperature of the tungsten filament is then computed using these physical parameters in Eq.~\ref{eq:stefan}. In the absence of a plasma, these values provide temperature in agreement with the electrical measurement of $\overline{T_W}$ within $30~$K.

\section{Current emitted by a heterogeneously heated filament}
\label{sec:empirical}

\begin{figure}[htbp]
    \begin{center}
        \includegraphics[width = \columnwidth]{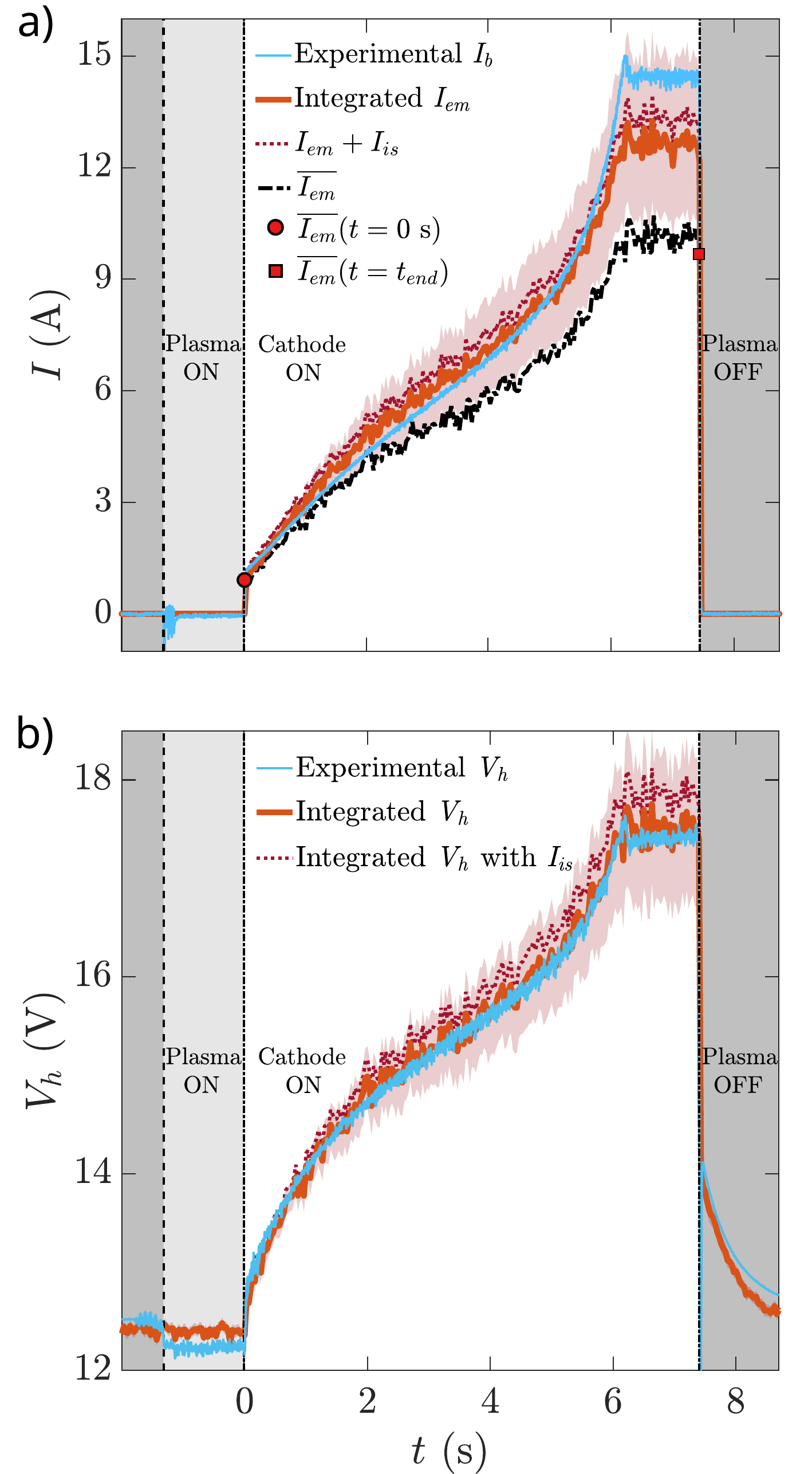}
        \caption{\textit{a)} Time evolution of the emitted current during a divergent regime ($I_h = 16.2~\mathrm{A}$, $V_b = –62~\mathrm{V}$): experimentally measured current $I_b$ (solid blue line),  integrated thermionic emission $I_{em}$ from temperature measurements (solid red line), sum of $I_{em}$ and ion current saturation $I_{is}$ (purple dotted line) and  effective thermionic emission $\overline{I_{em}}$ assuming a homogeneous temperature profile (solid black line). \textit{b)} Time evolution of the potential drop across the cathode:  experimentally measured voltage $V_h$ (solid blue line), integrated voltage taking into account temperature inhomogeneities (solid red line) and ion saturation current (purple dotted line) according to \eref{eq:u_h} (red). Red shaded areas correspond to errorbars for the solid red lines. See text for details.}
        \label{fig:analyse_ieth}    
    \end{center}
\end{figure}
In this section, we provide a detailed experimental characterization of a divergent regime, whose temporal evolution of the current $I_b$ is shown in the top panel of Fig.~\ref{fig:analyse_ieth}. The temporal evolution of the emitted current and voltage drop across the tungsten filament are reconstructed from spatially and temporally resolved measurements of the cathode temperature, in excellent agreement with the electrical measurements.

\subsection{Spatially and temporally resolved temperature measurements}

\begin{figure}[tbp]
    \centering
    \includegraphics[width = 0.48\textwidth]{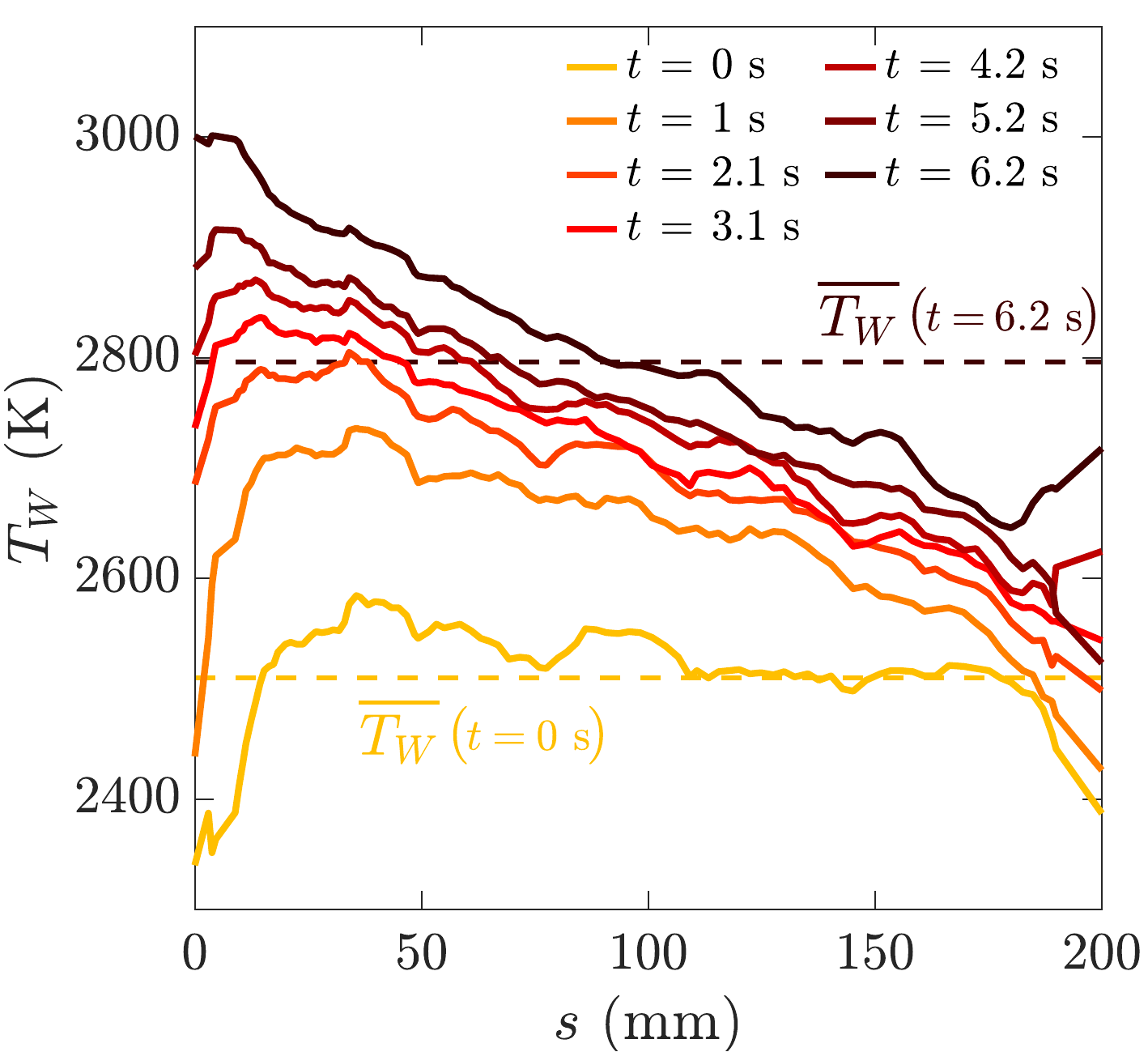}
    \caption{Temperature profile of the cathode along its curvilinear abscissa, from the center to the edge during the divergent regime ($I_h = 16.2~\mathrm{A}$, $V_b = –62~\mathrm{V}$) shown in Fig.~\ref{fig:analyse_ieth}.}
    \label{fig:Tw_exp}
\end{figure}

The spatial evolution of the tungsten filament temperature $T_W$ along the curvilinear abscissa $s$ is displayed in Fig.~\ref{fig:Tw_exp} at various time during the shot. 
As previously observed in Fig.~\ref{fig:photo_cath_plasma}, a nearly homogeneous temperature profile at $t=0$ quickly evolves to a strongly inhomogeneous profile, where the center of the spiral ($s\sim 0$) is hotter than the outer part of the spiral ($s\sim 200$~mm). A video of the evolution of the temperature profile is provided as a Supp. Mat. B. 
The balance between the various thermal processes is detailed in Sec.~\ref{sec:simulation}, but a rough sketch of the instability mechanism can be given, ignoring stabilizing processes. Thermionic electrons leave progressively the cathode from the center to the outer edge, leading to a decrease of the current flowing through the filament from the center to the outer edge and inducing an excess of Joule heating in the central part. The non-linear evolution of the Richardson current with $T_W$ then enhances the inhomogeneous heating, possibly leading to unstable divergent regimes, as in Fig.~\ref{fig:analyse_ieth}.
For the sake of illustration of the importance of spatially resolved temperature measurements, the value of the effective temperature $\overline{T_W}$ inferred from electrical resistance, is shown for $t = 0$ and $t = t_{\mathrm{end}} = 7.4~$s.  While an excellent agreement is observed for the average temperature, the electric current run-away can only be understood on the basis of temperature inhomogeneity.

\subsection{Estimate of thermionic emission from temperature measurements}
\label{sec:integ}

The total thermionic emission can be efficiently computed when plugging the spatio-temporal evolution of the cathode temperature $T_W$ into Eq.~\ref{eq:i_eth}. Assuming no space-charge limited effects, the total thermionic emission reads: 

\begin{eqnarray}
    I_{em}(t) & =  \int_0^{l_W} i_{em}(x,t) ~\mathrm{d}x\\
    & =\int_0^{l_W} A_G 2\pi r_W T_W(x,t)^2e^{-\displaystyle\frac{eW}{k_\mathrm{B}T_W(x,t)}} ~\mathrm{d}x,
    \label{eq:I_eth_tot}
\end{eqnarray}
where the notation $i_x$ denotes current by unit length. 
The time trace of $I_{em}(t)$ is shown in Fig.~\ref{fig:analyse_ieth}(a), solid red line. The shaded red area represents the error-bar for $I_{em}$ due to the error on the estimate of $T_W(s,t)$ \footnote{Since the 
calibration performed in the $2270~K$ - $2700~K$ range was extrapolated up to $3000~K$, an error $\Delta T = \pm 10~\mathrm{K}$ is applied for $T_W$ below 2700 K and $\Delta T = \pm 25~\mathrm{K}$ above 2700 K.}. A very good agreement is observed between the integrated thermionic emission $I_{em}(t)$ and the cathode current $I_b$ delivered by the bias power-supply, though $I_{em}(t)$ slightly  overestimates $I_b$ at the beginning of the shot and slightly underestimates $I_b$ during saturation.

The potential drop $V_h$ across the cathode is also computed from a similar integration, using the evolution of the  resistivity $\rho_W(T_W)$ with $T_W$.
The current $I_c(s,t)$ flowing in the cathode at position $s$ is the sum of the constant heating current $I_h$ and of the thermionic current emitted between position $s$ and the outer end of the cathode and reads: 
\begin{equation}
    I_c(s,t) = I_h + \int_s^{l_W} i_{b}(x,t)~\mathrm{d}x, 
    \label{eq:local_i}
\end{equation}

\noindent with $i_{b} = i_{em}$. The incorporation of the $10~\mathrm{m\Omega}$ contact resistance with the copper rods $R_{con}$ leads to:

\begin{eqnarray}
   V_h(t) = \int_0^{l_W}\rho_W\left(T_W(x,t)\right)&I_c(x,t)\frac{\mathrm{d}x}{\pi r_W^2} + \nonumber\\
   &R_{con}\left( I_h + \frac{I_b(t)}{2} \right)
    \label{eq:u_h}
\end{eqnarray}

The time traces of the experimental measurements and of the computation using spatially resolved cathode temperature profiles are shown in Fig.~\ref{fig:analyse_ieth}(b). The inhomogeneous heating of the cathode and the rise of $I_b$ result in a $42~\%$ increase of the cathode voltage drop  $V_h$ during the plasma shot, which is extremely well captured by the computation using the spatially and temporally resolved temperature measurement. Most of the increase in potential drop is due to the increase in $I_b$, as observed in the post-discharge regime where only the increase in filament resistance persists. 
A refinement might be incorporated by taking into account the ion saturation current $I_{is}$ in the current flowing through the negatively biased cathode. Note here that this is a small correction since, for the regimes reported in this article, $I_{em}/I_{is}$ lies in the range [5;16]. Due to the large primary injection from the cathode, the plasma density increases with thermionic emission. Following the experimental observations, a linear evolution of the plasma density (measured in the center of the plasma column) with the emitted current is considered here $n = 10^{18} + 3\times 10^{17}I_{em}$ (see Supp. Mat. C), and the Bohm velocity is assumed constant. We assume no plasma electron current flowing to the cathode since the  cathode potential is at least 10$T_e$ lower than  $\Phi_p$.
The small correction on the emitted current is displayed in Fig.~\ref{fig:analyse_ieth} (purple dotted line) for the cathode current and the voltage drop (for which $i_b(x,t)=i_{em}(x,t)+i_{is}$ in Eq.~\ref{eq:local_i}).

Finally, we stress the failure of using an effective temperature $\overline{T_W}$ from resistance measurement (which assumes an homogeneous temperature profile) for the prediction of the cathode current. While the experimental estimate of $\overline{T_W}$ is only possible at the beginning and at the end of the shot, it is possible to infer an effective average temperature from the spatio-temporal profiles, using the average electrical resistance value $\overline{R} = \displaystyle\int_0^{l_W}\rho_W\left(T_W(x,t)\right)\frac{\mathrm{d}x}{\pi r_W^2}$ in Eq.~\ref{eq:T}. An effective emitted current $\overline{I_{em}}$ is then computed using this value of the effective temperature in Eq.~\ref{eq:i_eth}. The results are shown in Fig.~\ref{fig:analyse_ieth} a), dotted black line. The experimental values using the experimentally measured values of $\overline{T_W}$ are shown as red-filled black symbols at  $t=0$ and $t=t_\mathrm{end}$. This method clearly underestimates the emitted current especially at strong emission, by up to 30\%.

\section{Thermionic emission modeling}
\label{sec:simulation}

\subsection{Cathode thermal budget and modeling}

The previous section demonstrated that the knowledge of the  spatiotemporal variations of the cathode temperature $T_W$ allows to precisely reconstruct the current $I_b$ drawn at the cathode. The goal of this section is to efficiently predict the temperature profile of the cathode by solving a local enthalpy budget equation leading to partial integro-differential equation for the temperature $T_W(s,t)$. We provide below a model that accurately predicts the spatio-temporal evolution of the cathode temperature during the divergent regime shown in Fig.~\ref{fig:analyse_ieth}, which thus applies both to stationary and unsteady regimes.
In the remaining of this Section, the notation $\dot{X}(s,t)$ refers to the time derivative per unit length $ds$ of the scalar $X$, \textit{i.e.} $\dot{X}(s,t) = \displaystyle\frac{\partial X(s,t)}{\partial t~ds}$. Further technical details about the simulation are provided in Supp. Mat. D.

The enthalpy budget of the cathode is  given by
\begin{equation}
    \dot{H} = \dot{Q}_\Omega + \dot{Q}_c  + \dot{Q}_i + \dot{Q}_{\sigma,in} - \dot{Q}_{\sigma,out}  - \dot{Q}_e,
    \label{eq:PDE}
\end{equation}
\noindent
where $H$ is the enthalpy of the cathode, and $\dot{Q}_X$ refer to powers per unit length. $\dot{Q}_\Omega$ is the ohmic heating term, $\dot{Q}_c$ the thermal conduction term, $\dot{Q}_i$ the heating term due to ion bombardment, $ \dot{Q}_{\sigma,in}$ the radiative incoming term, $\dot{Q}_{\sigma,out}$ the radiative outgoing term and $\dot{Q}_e$ the thermionic cooling term (or electron transpiration cooling term). Each of these terms are now discussed in details, and require the knowledge of the physical parameters of tungsten with temperature. The evolution with temperature were extracted from previous studies for the electrical resistivity $\rho_W(T_W)$~\cite{desai_electrical_1984, white_thermophysical_1997}, the specific heat $C_\mathrm{p}(T_W)$~\cite{white_thermophysical_1997}, the thermal conductivity $\lambda_W(T_W)$~\cite{white_thermophysical_1997}, and the total effective emissivity $\epsilon_\mathrm{eff}(T_W)$~\cite{Matsumoto1999HemisphericalTE} and are recalled in Supp. Mat. E.

\subsubsection{Detailed budget}

The evolution of the enthalpy is given by 
\begin{equation}
    \dot{H}(s,t) = C_\mathrm{p}(T_W) \rho \pi r_W^2 \frac{\partial~T_W(s,t)}{\partial~t},
    \label{eq:inertia}
\end{equation}
\noindent where $\rho$ is the volumic mass of tungsten.

The main heat source for the filament is ohmic heating, which depends upon the local current $I_c(s,t)$ flowing through the cathode at location $s$, given by Eq.~\ref{eq:local_i}: 
\begin{equation}
    \dot{Q}_\Omega(s,t) = \frac{\rho_W(T_W)}{\pi r_W^2} \left( I_h + \int_s^{l_W} i_{b}(x,t)~\mathrm{d}x \right)^2,
    \label{eq:ohmic}
\end{equation}
with $i_b(x,t) = i_{em}(x,t) + i_{is}$. Since $i_{em}$ is set by the temperature $T_W$ from Richardson's law, the ohmic heating term leads to an integral term in $T_W$.

Since the cathode is under vacuum, one can neglect the convection losses. Nonetheless heat diffuses along the filament via thermal conduction as:

\begin{equation}
    \dot{Q}_c(s,t) = \lambda_W(T_W) \pi r_W^2 \frac{\partial^2~T_W(s,t)}{\partial~s^2}
    \label{eq:conduct}
\end{equation}

Let us now discuss the radiative terms, and focus first on losses. The cathode is considered as a gray-body of hemispherical total emissivity $\epsilon_\mathrm{eff}(T_W)$, which is the total effective emissivity of tungsten along the complete spectra of radiation, and the emitted radiative power per unit length reads:

\begin{equation}
    \dot{Q}_{\sigma,out}(s,t) = –\sigma 2\pi r \epsilon_\mathrm{eff}(T_W)T_W(s,t)^4
    \label{eq:sigma_loss}
\end{equation}

\begin{figure}[tbp]
    \centering
    \includegraphics[width = 0.3\textwidth]{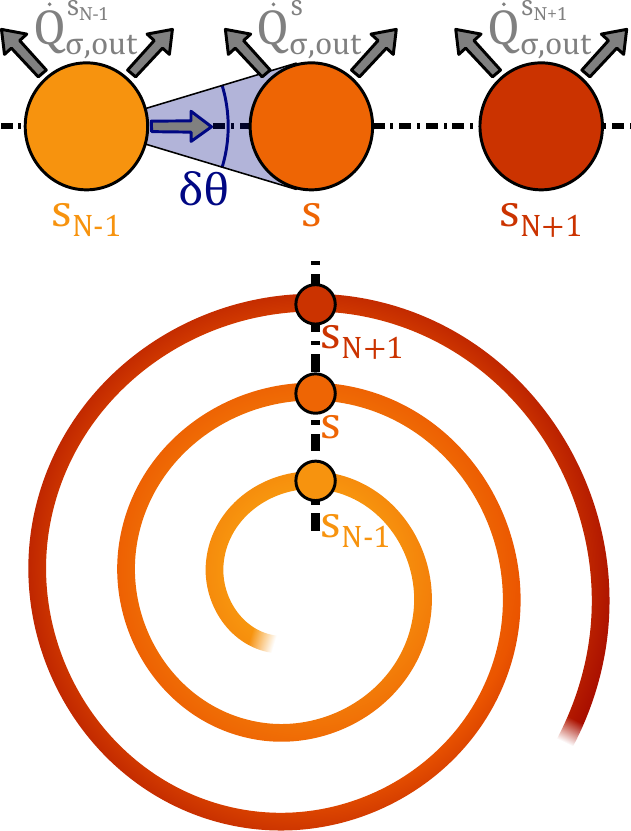}
    \caption{Cross-sectional (top) and upper (bottom)  views of the cathode. The solid angle $\delta \theta$ is shown in light blue and gives an estimation of the parameter $\alpha$.}
    \label{fig:alpha_expl}
\end{figure}

The incoming radiative flux is mainly due to self-heating by the radiation from neighbouring part of the filament. The radiative inward flux from the environment and the low-density plasma are neglected (the vessel and the argon neutrals are close to room temperature). The winding of the spiral leads to an inward flux at a given location $s$ (on turn $N$), emitted by the cathode at location $s_{N-1}$ on turns $N-1$ and location $s_{N+1}$ on turn $N+1$, as sketched in Fig.~\ref{fig:alpha_expl}. A fraction $\alpha$ of the
radiative power per unit length $\dot{Q}_{\sigma,out}(s_{N-1},t)$ emitted at location $s_{N-1}$ is intercepted by the filament at location $s_N$. The value of $\alpha$ is approximately given by the value of the solid angle $\delta \theta$ seen by each filament, as $\alpha \simeq \delta \theta / 2\pi \simeq 0.1$. Assuming that the absorbance of the filament at $T_W$ is $\epsilon_\mathrm{eff}(T_W)$, the radiative inward term at position $s$ reads:

\begin{eqnarray}
   \dot{Q}_{\sigma,in}(s,t) = &\alpha\epsilon_\mathrm{eff}(T_W) \nonumber \\
   &    \left(\dot{Q}_{\sigma,out}(s_{N-1},t)+ \dot{Q}_{\sigma,out}(s_{N+1},t) \right)
    \label{eq:sigma_gain}
\end{eqnarray}

The exact value of the parameter $\alpha$ was computed from the measurement of the cathode spiral in the absence of a plasma, for heating currents ranging from 12 to 18.5 A, resulting in a value $\alpha = 0.15$ close to the 0.1 estimate (see Supp. Mat. F for details).

The last heating source for the cathode is due to the plasma-cathode interaction, namely ion bombardment, since interactions with plasma electrons and neutrals will be ignored. 
No secondary-electron emission is taken into account as above-mentioned~\cite{konuma_film_1992, tolias_secondary_2014}. The energy of ions reaching the cathode is the sum of the potential energy $e\left(E_i-W\right)$ due to the ion-electron recombination, with $E_i$ the ionization energy in eV, and the kinetic energy $\beta e\left(V_c(s,t)~–~\Phi_p(t)\right)$ \cite{kersten_energy_2001} where $\beta$ is a numerical factor to account for dissipative phenomena during the energy transfer to the lattice, which is expected to lie between 0.5 and 1~\cite{kersten_energy_1994}, and $V_c$ the local potential of the tungsten filament. The ion-bombardment term thus reads: 
\begin{equation}
    \dot{Q}_i = \frac{i_{is}}{e} e\left[E_i~–~W + \beta\left(V_c(s,t)~–~\Phi_p(t)\right)\right]
    \label{eq:ion_heat}
\end{equation}
The only adjustable parameter is $\beta$, chosen such that the total current emitted by the cathode computed according to the model matches the experiments (see Fig.~\ref{fig:Tw_exp_vs_simu}).

Thermionic electrons extracts heat from the material as they leave the cathode~\cite{herring_thermionic_1949, Hanquist_2017}. Assuming a Maxwell energy distribution at temperature $T_W$ for the emitted electrons, the cooling term reads:

\begin{equation}
    \dot{Q}_e = –\frac{i_{em}}{e}\left( eW + 2k_\mathrm{B}T_W\right)
    \label{eq:thermion_cool}
\end{equation}

\noindent where $i_{em}$ is the thermionic current per unit length.

\subsubsection{Boundary conditions}
\label{sec:boundary}

Boundary conditions are set to prescribe heat fluxes at both ends. The cathode is clamped to copper rods of thermal conductivity $\lambda_{Cu}$, length $l_{Cu}$, radius $r_{Cu}$ which are kept at ambient temperature at the other end. Thermal conduction within the large copper rods being the only flux term, this imposes the spatial derivatives of temperature $T_W$ at the boundaries:

\begin{equation}
    \cases{
    \left. \frac{\partial~T_W}{\partial~s}\right|_{s = 0} = \frac{\lambda_{Cu} r_{Cu}^2}{\lambda_W(T_W(0,t)) r_W^2}\frac{T_W(0,t)~–~T_{amb}}{l_{Cu}}\\
    \left. \frac{\partial~T_W}{\partial~s}\right|_{s = l_W} = –\frac{\lambda_{Cu} r_{Cu}^2}{\lambda_W(T_W(l_W,t)) r_W^2}\frac{T_W(l_W,t)~–~T_{amb}}{l_{Cu}}
    }
    \label{eq:bc}
\end{equation}

The poor thermal contact between tungsten and copper leads to set $\lambda_{Cu} = 40~\mathrm{W/(m~K)}$, one order of magnitude below the common values for $\lambda_{Cu}$~\cite{white_thermophysical_1997}. The weak influence of this parameter is discussed in Supp. Mat. G.

\subsection{Computation of the cathode current}

The temporal evolution of the cathode temperature profile $T_W(s,t)$ is computed from the numerical integration of the integro-differential equation for $T_W$ (see Supp. Mat. D and J). Assuming no space-charge limitation, the cathode current is the sum of the emitted current computed using the spatio-temporal evolution of $T_W$ in Eq.~\ref{eq:I_eth_tot} and of the ion saturation current $I_{is}$. The space-charged limited solution is implemented according to the model from Ye and Takamura \cite{ye_effect_2000} and bounds the thermionic current in case of strong emission and weak bias.

\subsection{Modeling the influence of the emitted current on plasma parameters}
\label{sec:plasma}

As shown by the experimental results displayed in Fig.~\ref{fig:div_regime}, the dynamical evolution of the cathode current is very sensitive to the details of the plasma-cathode interactions. Hence the prediction of the cathode temperature profile (and thus of the cathode current) depends upon the evolution of the plasma density and plasma potential with the cathode current. It is important to provide a precise picture of the evolution of plasma parameters with the cathode current.

Let us first focus on the plasma potential. Despite the recent development of analytical models for the evolution of the plasma potential in the presence of emitted current~\cite{trotabas_trade-off_2022, poulos_model_2019}, the limitations of these models do not allow direct comparison with experimental observations at very large current emission~\cite{pagaud_sub2023}. We chose to model the evolution of $\Phi_p$ following a power law evolution as :
\begin{equation}
    \Phi_p = \Phi_{p,0} + \gamma I_{b}^{1/p}, 
    \label{eq:Vp}
\end{equation}
where $\Phi_{p,0} = -1$~V is the plasma potential at zero thermionic emission and $\gamma=-6.6$ and $p=3$ are empirical constants. This evolution is computed from experimental measurement of the plasma potential during a divergent regime using an emissive probe in the center of the plasma column, for which the plasma parameters are nearly invariant along the axial direction (see Supp. Mat. H).

The evolution of the plasma density with the cathode current could also be modeled empirically, as was done in the previous section with a linear evolution as a function of the cathode current (see Supp. Mat. C). However, we decided to compute the plasma density at each time step following a simplified power balance~\cite{lieberman,ChabertBraithwaite}. Assuming that the additional power injected by the cathode is lost from recombination at the walls, the plasma density increase $\Delta n$ due to thermionic electrons at the center of the plasma column reads  $\Delta n = 3P_{cath}/\pi R^2\sqrt{eT_e/m_i}e\mathcal{E}_\mathrm{T}$, with $\mathcal{E}_\mathrm{T}$ the total energy lost per electron–ion pair. The plasma density increase is then used to compute the ion saturation current, assuming that the electron temperatures remains constant, both for the computation of the ion bombardment term and of the ion saturation current drawn at the cathode. The simplified power balance implemented here fairly well reproduces the global ion saturation flux measured experimentally (see Supp. Mat. C).

\subsection{Simulation of the temporal dynamics of diverging regimes}

\begin{figure}
    \centering
    \includegraphics[width = 0.48\textwidth]{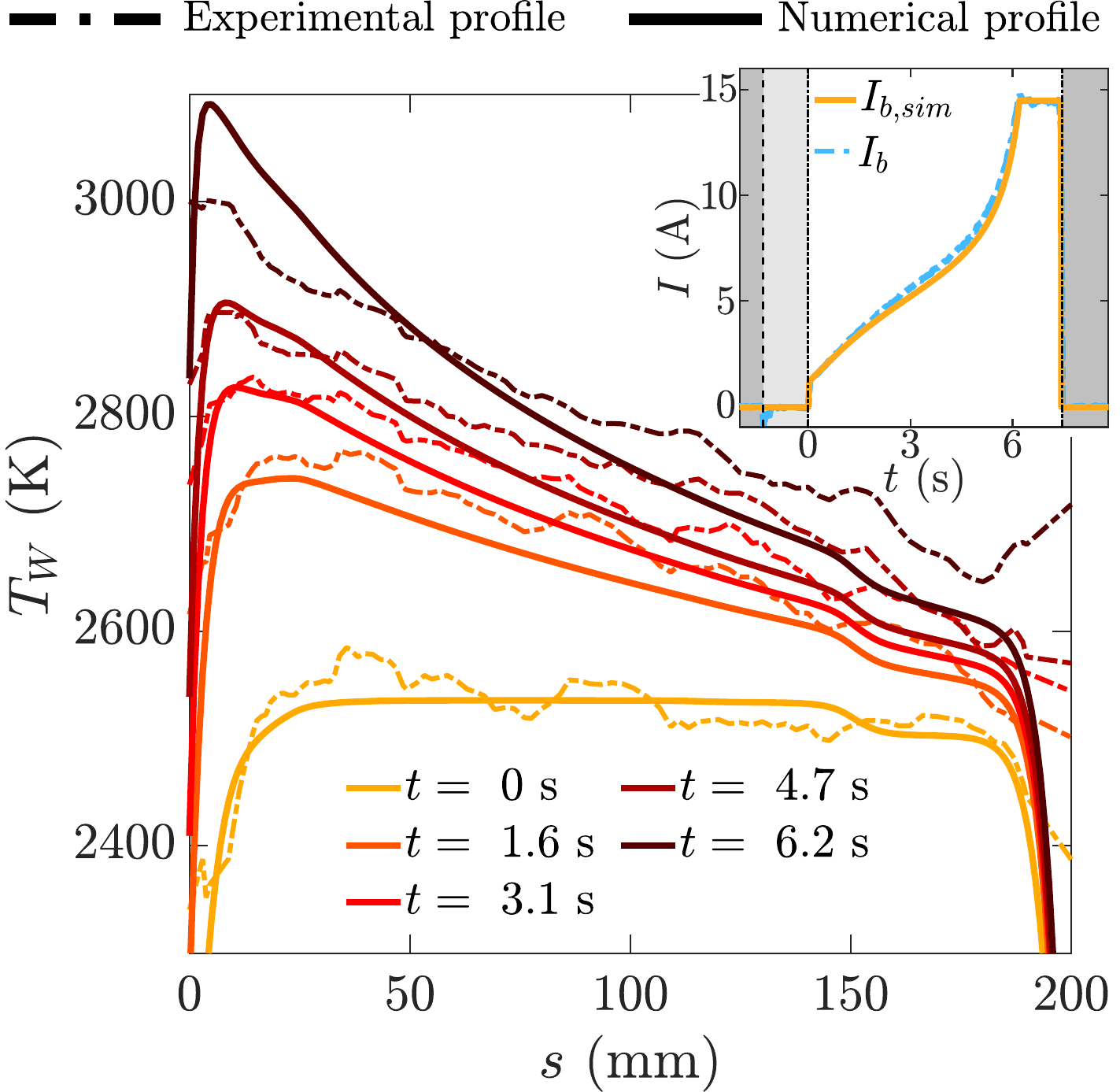}
    \caption{Comparison of experimental and numerical $T_W$ during a divergent regime ($I_h = 16.2~\mathrm{A}$, $V_b = –62~\mathrm{V}$). Dashed lines are for experimental profiles, solid lines are for numerical profiles. The same colors correspond to identical timestamps. Insert represents the current $I_b$ over time obtained experimentally and numerically.} 
    \label{fig:Tw_exp_vs_simu}
\end{figure}

Finally, the models described above are numerically solved for the conditions of the dynamics shown in Fig.~\ref{fig:analyse_ieth} and \ref{fig:Tw_exp}. The parameter $\beta$ (the ratio of energy transfer considered for the ion bombardment term) is the only free parameter of the model, whose value is set to match the experimental temperature profiles. This results in the temporal evolution of the spatial profiles of the cathode temperature shown in Fig.~\ref{fig:Tw_exp_vs_simu}, which very accurately reproduces the experimental measurements, with $\beta = 1.01$. The inset of Fig.~\ref{fig:Tw_exp_vs_simu} shows that the simulation of the total cathode current is in excellent agreement with the experimental value. Deeper insight on the influence of $\beta$ is given in Supp. Mat. I. A video showing the temporal evolution of the experimental and numerical profiles is provided in Supp. Mat. B. The model very accurately reproduces the experimental profiles during the first part of the shot. Then, when the regime diverges rapidly, the numerical profiles are slightly warmer than the experimental profiles on the first turns of the spiral and slightly cooler on the outer turns. We stress that the discrepancy between the profiles at the centre of the cathode might originate in an underestimated experimental temperature due to the pyrometer calibration, as mentioned in section \ref{sec:integ}. The limitations of the model are further discussed in Sec.~\ref{sec:limit}. The influence of the influence of the two parameters arbitrarily set to reproduce the experimental temperature profiles, namely $\lambda_{Cu}$ and of $\beta$, are further discussed in Supp Mat, and show that $\lambda_{Cu}$ has a rather weak influence on the simulated profiles, while the value $\beta$ significantly alters the heating of the cathode due to  plasma-cathode interactions and is the most important player in the model.

\subsection{Limits of the model}\label{sec:limit}

While the model reproduces the experimental data with high fidelity, we now discuss some limitations.The first refinement would concern including heating from electron bombardment, which would be important when $\Phi_p~–~(V_b + V_h)$ is lower than $\sim 3~T_e$. In addition, the cathode properties might change during a divergent regime. Tungsten sputtering is not taken into account, even though scanning electron microscopy has shown that the cathode radius $r_W$ shrinks by around 2 \% after a few hundred plasma shots. Tungsten evaporation would add a new cooling term in the energy budget equation and the decrease of $r_W$ would modify the energy budget equation. Finally the work function $W$ is set constant while a modification of $10~\mathrm{meV}$ would be sufficient to affect the onset of diverging regimes. These limits could explain the slight discrepancies between the experimental and simulation profiles of Fig.~\ref{fig:Tw_exp_vs_simu}. The discrepancy at the center could be attributed to two reasons. The experimental temperature might be underestimated due to the pyrometer calibration, as mentioned in section \ref{sec:integ}. The simulation profile could, on the other hand, be overestimated since the cooling evaporation of tungsten, which grows with thermionic emission~\cite{goebel_lab6_2007}, has been neglected. At the outer end of the cathode, the lower values of the cathode temperature in the simulation might be due to the absence of electron bombardment in simulations. Indeed, the outer part of the filament reaches higher potentials, leading to more electron bombardment and thus heating. 

Finally, note that the precise tuning of parameters could slightly depend upon the base plasma parameters $n_0$ and $T_{e,0}$ and thus upon base pressure, magnetic field, RF power and geometry of the plasma device.

\section{Operation of a highly emissive cathode immersed in plasma: insights from modeling}
\label{sec:operation}

\subsection{Prediction of the operating parameters and of  stable regime limits}

\begin{figure}
    \centering
    \includegraphics[width = 0.95\columnwidth]{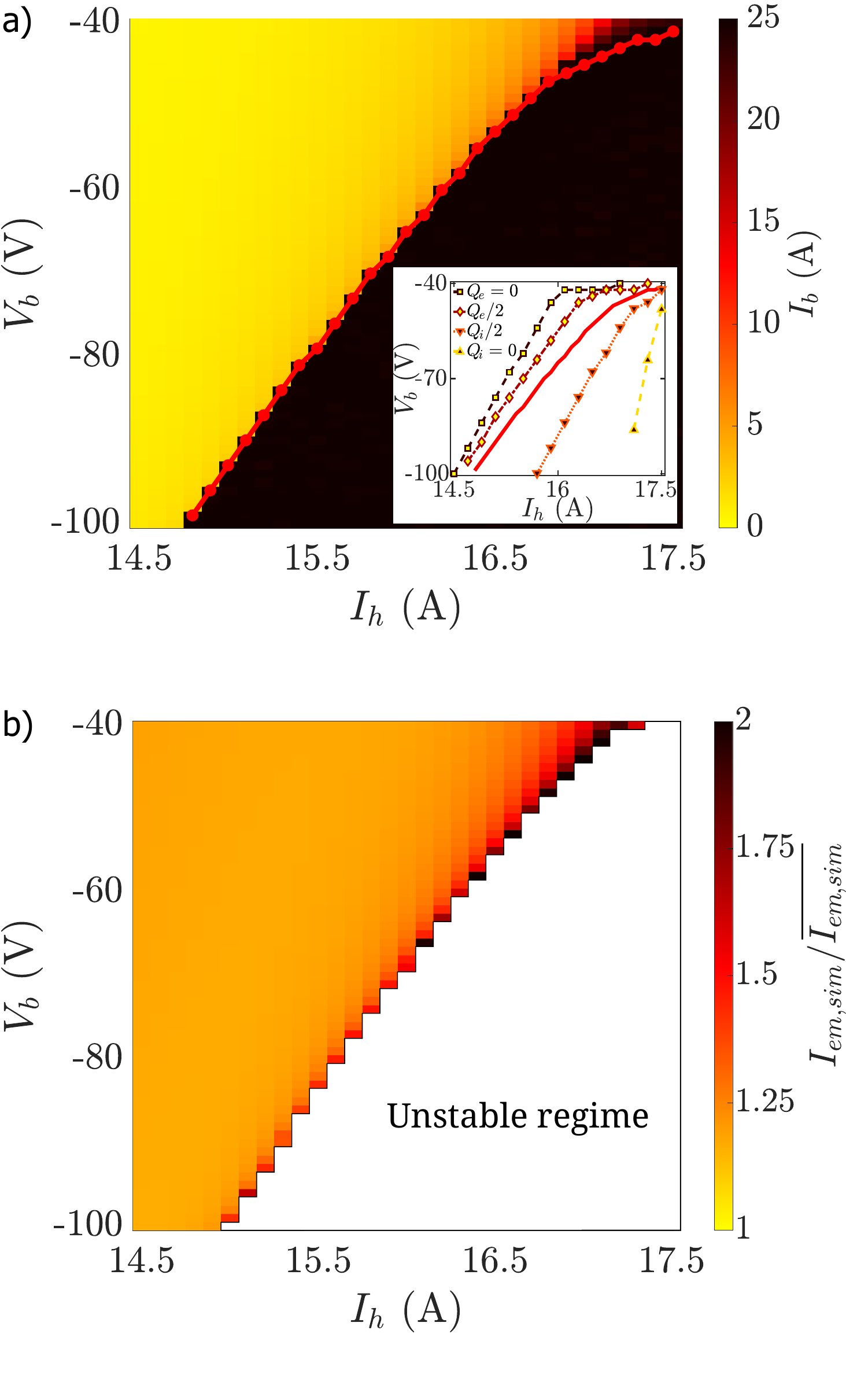}
    \caption{\textit{a)} $I_b$ as a function of experimental parameters $V_b$ and $I_h$ after $30~$s of simulation. Simulations are bounded at $25~A$ maximum. The red line splits the diagram in two domains whether $I_b = 25~\mathrm{A}$ is reached or not. Inset: frontiers between the stable and divergent regimes for different values of $\dot{Q}_i$ and $\dot{Q}_e$. \textit{b)} Corresponding ratio $I_{em,sim}/I_{em,elec}$ after $5~$s of simulation.}
    \label{fig:phase}
\end{figure}

The model described in the previous section allows to predict whether a steady-state regime or a time-divergent regime is reached for cathode operation. For given values of the background plasma parameters, a set of simulations were run for heating current $I_h\in \left[ 14.5~\mathrm{A} ; 17.5~\mathrm{A}\right]$ and bias voltage $V_b\in\left[ –100~\mathrm{V} ; –45~\mathrm{V}\right]$. The values of the cathode current $I_b$ obtained after 30-seconds simulations are reported in Fig.~\ref{fig:phase}. Diverging regimes, similar to the regimes previously described,  are defined when the cathode exceeds 25 A, lie below the red curve in Fig.~\ref{fig:phase}, \textit{i.e.} the black portion of the parameter map, at high heating current and high voltage bias. Stable regimes at moderate to low emission are observed for low heating current or low voltage bias. The features highlighted in Fig.~\ref{fig:div_regime} are thus correctly captured by the model. A high emission would cause severe damage to the cathode due to tungsten sputtering, and is therefore undesirable. Note that  stable regimes at strong emission are observed for moderate bias and strong heating in the upper right part of the parameter space ($I_b \in \left[15~\mathrm{A} ; 25~\mathrm{A}\right]$); this should be interpreted with caution as $I_e$ can no longer be ignored since $(V_b + V_h-\Phi_p)/T_e$ is below 3. The inset highlights the sensitivity to  $\dot{Q}_i$ and $\dot{Q}_e$ and displays the frontiers between stable and divergent regimes for four cases: $\dot{Q}_i = 0$ (yellow up triangles), $\dot{Q}_i/2$ (orange down triangles), $\dot{Q}_e = 0$ (black squares) and $\dot{Q}_e/2$ (brown diamonds). One can see notably the stabilizing effect of thermionic cooling and most importantly the prominent role of ion bombardment in the dependence on $V_b$ of the runaway behaviour.

The experimental data confirm the extreme sensitivity with the control parameters, and the existence of divergent regimes around ($I_h = –16.2~\mathrm{A}$, $V_b = –62~\mathrm{V}$), as well as the non-existence of a divergent regime at $V_b > –40~\mathrm{V}$ for any value of $I_h$. Sharp transitions from stable to unstable regimes at $V_b < –80~\mathrm{V}$ and $I_h \sim 15.5~\mathrm{A}$ have also been observed.

\subsection{Correction of current emission computed using the effective cathode temperature from electrical measurements}

As already mentioned, thermionic electron emission is clearly underestimated when using the effective cathode temperature $\overline{T_W}$ from electrical measurements (see Fig.~\ref{fig:analyse_ieth}). However, the ease with which the effective cathode temperature can be derived from the overall electrical resistivity makes it very attractive. Using the complete thermal modeling of the cathode presented in Sec.~\ref{sec:simulation}, we are able to compute a correction factor for the thermionic emission computed as $I_{em, sim}/\overline{I_{em, sim}}$, as shown in the bottom panel of Fig.~\ref{fig:phase}. The correction factor ranges between 1.2 and 1.5 for most of the operational parameters, but may reach values up to 2 for large emitted current. The use of this correction factor is extremely important for the estimate of the the Richardson current to predict, for instance, how current injection affects the potential profile~\cite{trotabas_trade-off_2022,pagaud_sub2023}

\subsection{Contributions of heat transfer mechanisms}

\begin{figure*}
    \centering
    \includegraphics[width = 0.95\textwidth]{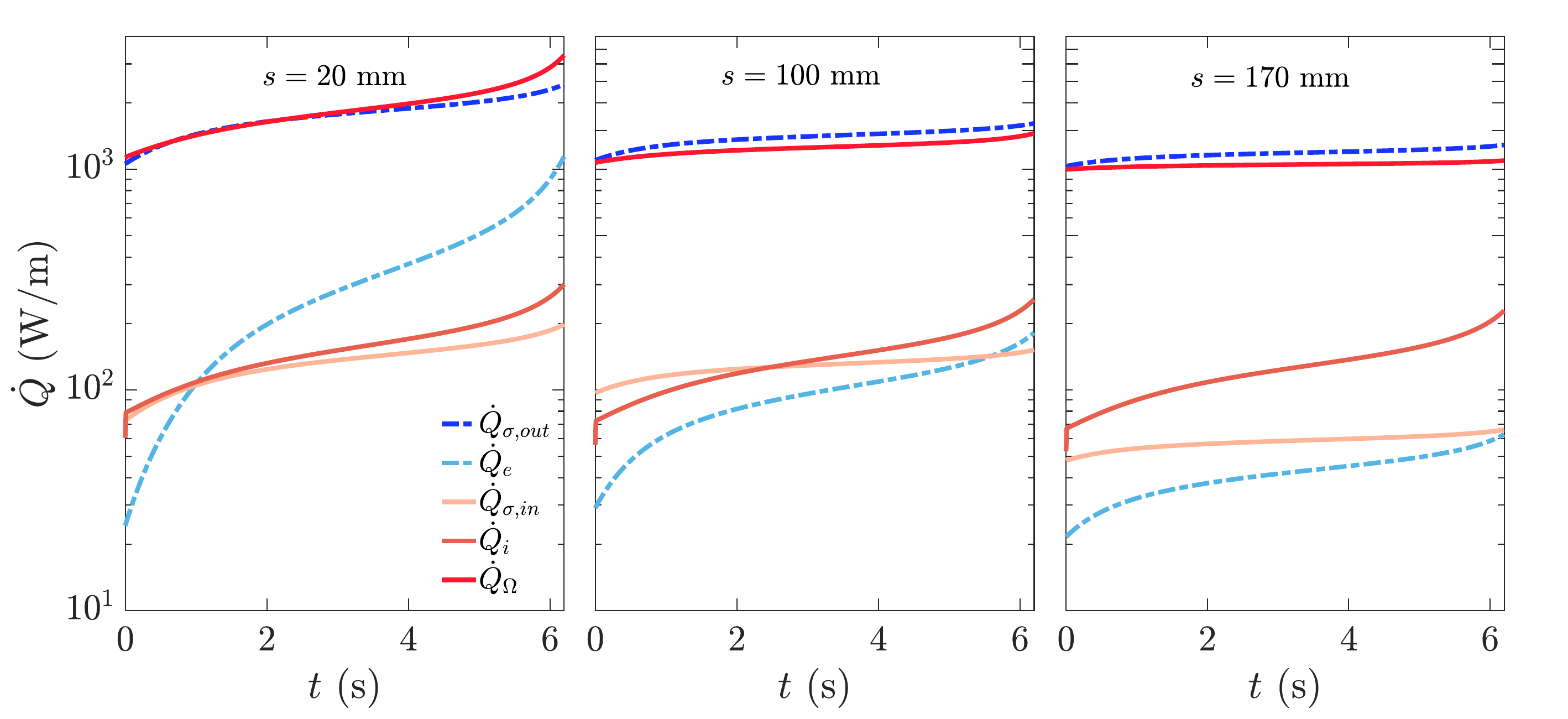}
    \caption{Power per meter provided to the filament over time at $s = 20~$mm, $s = 100~$mm and $s = 170~$mm. Red solid lines are the heating terms, blue dashed lines are the cooling terms. Conduction is not represented here.}
    \label{fig:bilan}
\end{figure*}

The simulation enables to compute the various heat sources and sinks, and to assess the importance of each of them separately. Hence the heating mechanisms at play are displayed in Fig.~\ref{fig:bilan} at three different locations along the cathode over time (conduction is subdominant, except near the extremities). A video illustrating time evolution of the various terms is available in Supp. Mat. B.

Heating mechanisms are represented by solid red lines, cooling mechanisms by dashed blue lines. One can see that $\dot{Q}_\Omega$ and $\dot{Q}_{\sigma,out}$ are an order of magnitude greater than the other terms and account for the main dynamics of $T_W$. On the one hand, $\dot{Q}_i$ and $\dot{Q}_{\sigma,in}$ are rather homogeneous heating terms along the cathode length. The latter presents almost no variations over time, illustrating its weak effect on the runaway behaviour. $\dot{Q}_i$ causes a supplementary heating at $t = 0~$s that depends on $V_b$ and $n$, which might be enough to trigger the runaway behaviour as the density of the plasma increases with emission. The inset of Fig.~\ref{fig:phase} highlights the importance of ion bombardment on the influence of $V_b$ witnessed in Fig.~\ref{fig:div_regime}. 
On the other hand, $\dot{Q}_e$ being proportional to $T_W^2 \exp\left(-eW/k_\mathrm{B}T_W\right)$, its heterogeneity is important and it provides an essential stabilizing effect, especially at the center of the cathode (around $s = 20~$mm), where the cathode temperature is the highest. This is particularly important at late times.

On the other hand, the temperature inhomogeneity provides an essential stabilizing effect through $\dot{Q}_e$, which is proportional to $T_W^2 \exp\left(-eW/k_\mathrm{B}T_W\right)$. This is particularly observed at the center of the cathode (around $s = 20~$mm), where the cathode temperature is the highest, especially at late times.

The runaway behaviour can thus be analyzed according to the following unstable ingredients:
\begin{itemize}[label = $\bullet$]
    \item Thermionic emission causes more current to flow through the cathode, resulting in an enhanced heating at the cathode center (biased at the lowest potential $V_b$), which causes a higher thermionic emission and a heterogeneous temperature profile. This effect is more pronounced with a thinner filament.
    \item The density of the plasma grows with thermionic emission, resulting in a stronger ion bombardment and thus more heating. This is also true to a smaller extent with an increased electron temperature, since ion bombardment is s proportional to $n_e \sqrt{T_e}$. This effect increases with a thicker filament as ion bombardment is proportional to the cathode surface. 
\end{itemize}

On the other hand, the stabilizing feedback are the following:
\begin{itemize}[label = $\bullet$]
    \item Dissipative radiation is the main source of heat losses for the cathode. As it grows with $T_W^4$, a rise in $T_W$ results in a much stronger dissipation.
    \item Thermionic cooling may become an important cooling term at high emission, since this stabilizing effect is proportional to the thermionic current. 
    \item The plasma potential decreases with thermionic emission, resulting in weaker ion bombardment. Even though it is a minor effect, it can modify slightly the frontier of the stability regime.
    \item Space-charge limited regimes may also reduce emission when $(V_b-\Phi_p)/T_e$ is of order unity. For instance, this prevents the divergence of the emitted current beyond $V_b = -40~\mathrm{V}$ as $I_b$ saturates for any value of $I_h$
\end{itemize}


\section{Conclusion}

In conclusion, the regime of operation, and in particular the cathode current of a highly emissive cathode in a pre-existing plasma is shown to be accurately predicted from the knowledge of the temperature profile along the cathode. Spatially and temporally resolved temperature profiles have been obtained using an intensity comparison pyrometer, and the computation of the emitted current using Richardson's law is in excellent agreement with the measurements. The experimental results are also predicted as the solution of a detailed thermal balance, resulting in an integro-differential equation for the temperature field. 
It was found that, when immersed in a high density argon plasma column, the cathode undergoes strong temperature heterogeneities mainly due to the inhomogeneous ohmic heating from thermionic current. This causes the emission current not only to be driven by the initial temperature of the cathode, but also by the bias of the cathode and the plasma properties. The thermal model shows that the regime of operation is highly sensitive to the cathode heating from ion bombardment.
This model provides insight in the physics at stake for hot emissive metal filaments, as well as a predictive tool useful to implement safely such an experimental configuration. 

The deep understanding of this simple cathode design opens the path to a wider use of cathodes as control tools in pre-existing plasmas at a minor cost. However this article found an operational regime that limits the thermionic current in order to avoid damaging the tungsten filament. 
Though this design is an interesting alternative to costly and complex oxide cathodes at low emission, systematic studies at very large emission require the development of oxide cathodes, such as $\mathrm{LaB_6}$ cathodes.

\section*{Acknowledgements}
The authors acknowledge fruitful discussions with Renaud Gueroult.

\section*{Author Declarations}
\subsection*{Conflict of Interest}
The authors have no conflicts to disclose.

\subsection*{Data availaility}
The data that support the findings of this study are available from the corresponding author upon reasonable request. The numerical routines of the model described in Sec.~\ref{sec:simulation} are freely available on a github repository \url{https://github.com/FrancisPagaud/Emissive-Cathode-Model.git}

\subsection*{Author Contributions}
Francis Pagaud: Conceptualization (lead); Data curation (lead); Formal analysis (lead); Investigation (equal); Resources (supporting); Software (lead); Writing – original draft (lead); Writing –  editing (equal).\\
Vincent Dolique: Conceptualization (supporting) 
; Investigation (supporting); Resources (supporting); 
Writing –  editing (supporting).\\
Nicolas Claire: Investigation (supporting); Resources (supporting);  Writing –  editing (supporting).\\
Nicolas Plihon:
Conceptualization (supporting); Data curation (supporting); Formal analysis (supporting); Funding acquisition (lead); Investigation
(equal); Resources (lead); Software (supporting); Writing – original draft (supporting); Writing – editing (equal); Supervision (lead).

\providecommand{\newblock}{}

\clearpage
\newpage
\begin{center}{\Large  Supplementary data}
\end{center}


\section{Transmittance of the glass window}
\label{sec:transmit}

The glass window separating the cathode and the pyrometer absorbs partially the light at $650~$nm. Its transmittance has been quantified from relative measurements following the following protocol:

\begin{itemize}
    \item[1.] The cathode is put under vacuum in front of an unspecified borosilicate window and heated at a fixed heating current $I_h$. Its apparent temperature $T_W$ is measured at an apparent given emissivity $\epsilon_a$, arbitrarily set for the pyrometer (which takes into account the real emissivity of the cathode and the transmittance of the unspecified window).
    \item[2.] Then the borosilicate window used for the experiment presented in section 3 is added outside of the vacuum chamber between the pyrometer and the cathode. The emissivity paramater of the pyrometer r is adjusted to $\epsilon'$ in order to recover a measured temperature equal to $T_W$, obtained at step 1. This relative measurement provides the transmittance $\tau_\mathrm{g}$ of the borosilicate window as the ratio $\epsilon'/\epsilon_a$.
    \item[3.] The experiment is repeated for various values of $I_h$ and $\epsilon_a$ to ensure reproducibility.
\end{itemize}

\begin{figure}[htbp]
    \centering
    \includegraphics[width = 0.48\textwidth]{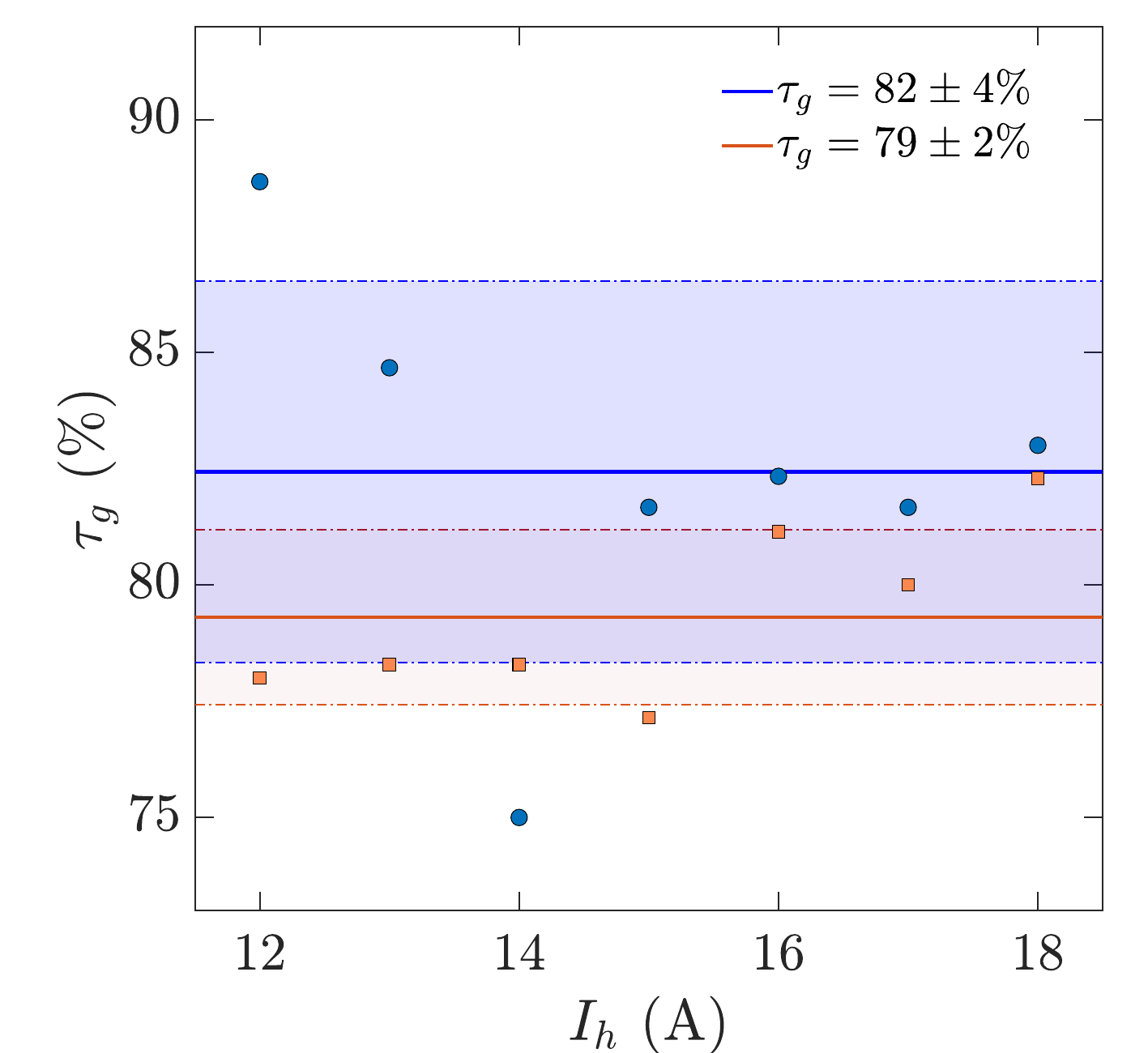}
    \caption{Transmittance of the glass window between the pyrometer and the cathode for $\lambda = 650~\mathrm{nm}$ for two different scans in $I_h$. }
    \label{fig:transmit}
\end{figure}

The real emissivity $\epsilon$ used in the experiment in section 3 is then $\epsilon = \epsilon_W \tau_\mathrm{g}$, with $\epsilon_W$ the tungsten emissivity. The correction factor $\tau_\mathrm{g}$ is displayed in \fref{fig:transmit} for two different initial values of $\epsilon_a$ as a function of $I_h$. Dark blue (respectively light orange) solid lines represent the mean values of the dark circle (resp. light square) markers, pale areas represent the standard deviation for each dataset. The final value of the glass transmittance has been assessed at 80 $\pm 4~\%$.

\begin{figure*}[htbp]
    \centering
    \includegraphics[width = 0.80\textwidth]{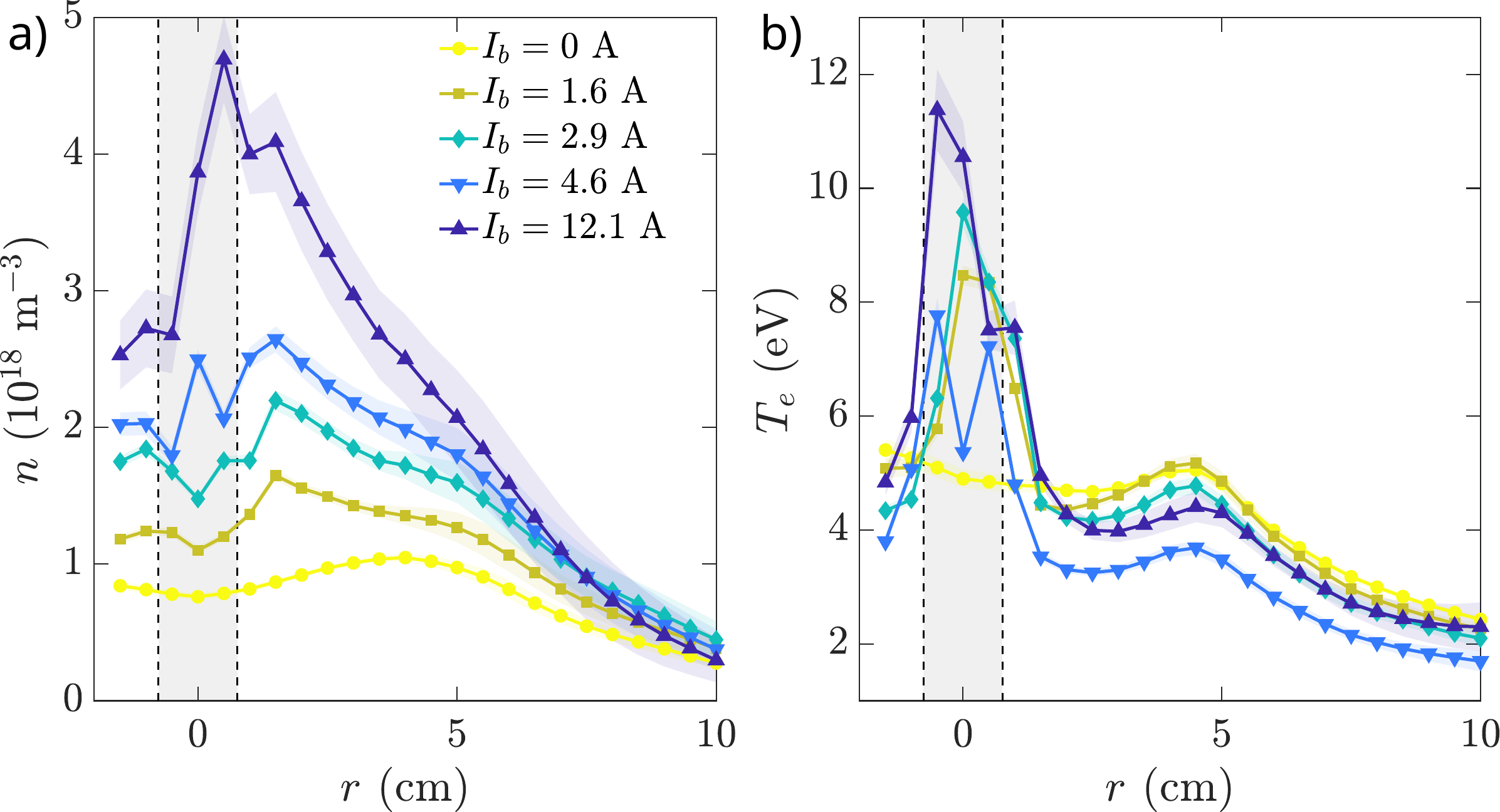}
    \caption{Radial evolution of \textit{a)} the plasma density and \textit{b)} the electron temperature for various currents $I_b$. The gray areas represent the location of the cathode.}
    \label{fig:n_r}
\end{figure*}

\section{Videos of operation, measurements and simulation}
\label{sec:video}

\verb|Video_1.mp4| shows the effect of thermionic emission on the plasma column. It is a side view of the inside of the plasma chamber at 1 mTorr, 170 G and 1 kW. A drawing of the setup is found at the beginning: the RF source is at the left end, while the cathode faces the plasma at the other end of the chamber. First the plasma is ignited using the RF source on the left side, resulting in a 10-cm wide plasma column while the cathode is hot and floating. Then the cathode is negatively biased and emits primary electrons from the right end, creating a denser 2-cm wide plasma core. The strong emission at the cathode is sustained for a few seconds, and finally the power of the RF source is shut down.

\verb|Video_2.mp4| is a recording of the light intensity through the pyrometer for the experimental conditions presented in section 3. The cathode is observed through the pyrometer, which filters the light at $650~$nm. Initially, the reference gray body of the pyrometer ($\Lambda$ shape) is brighter than the cathode. The plasma is first ignited and the cathode is negatively biased, resulting in a strong thermionic emission and a progressive cathode heating, finally ending in a divergent regime. The heterogeneous intensity profile is highlighted here. An homogeneous profile is quickly recovered once the plasma is turned off.

\verb|Video_3.mp4| shows simultaneously the raw pyrometer measurements (left panel), the temperature profile along the cathode extracted from the raw pyrometer measurements (central panel) and the emitted current (right panel).

\verb|Video_4.mp4| shows the temporal evolution of the measured temperature profile and the numerical resolution for $I_h = 16.2~$A, $V_b =~–62~$V and $\beta = 1.01$.

\verb|Video_5.mp4| shows the temporal evolution of the spatial profiles of the heating and cooling terms (powers per unit length) for the numerical solution. The red positive terms are the source terms while the blue negative terms are the sink terms. Conduction is not represented here but represents only a significant part at the 15-mm extremities. The black dashed-line is the sum of all the terms, conduction included.

\section{Plasma density increase with thermionic emission}
\label{sec:n_ib}

The plasma density increases with the cathode current $I_b$. The evolution of the plasma density with the emitted current is taken into account in Section 4 for the computation of the contribution of the ion saturation current in the total cathode current. It is also taken into account in Section 5 for the thermal modeling, including the power deposition from the emissive cathode. 

The radial evolution of the plasma density and the electron temperature are shown in \fref{fig:n_r} for $V_b = ~–60~$V for various values of the total cathode current $I_b$.

\begin{figure}[htbp]
    \centering
    \includegraphics[width = 0.48\textwidth]{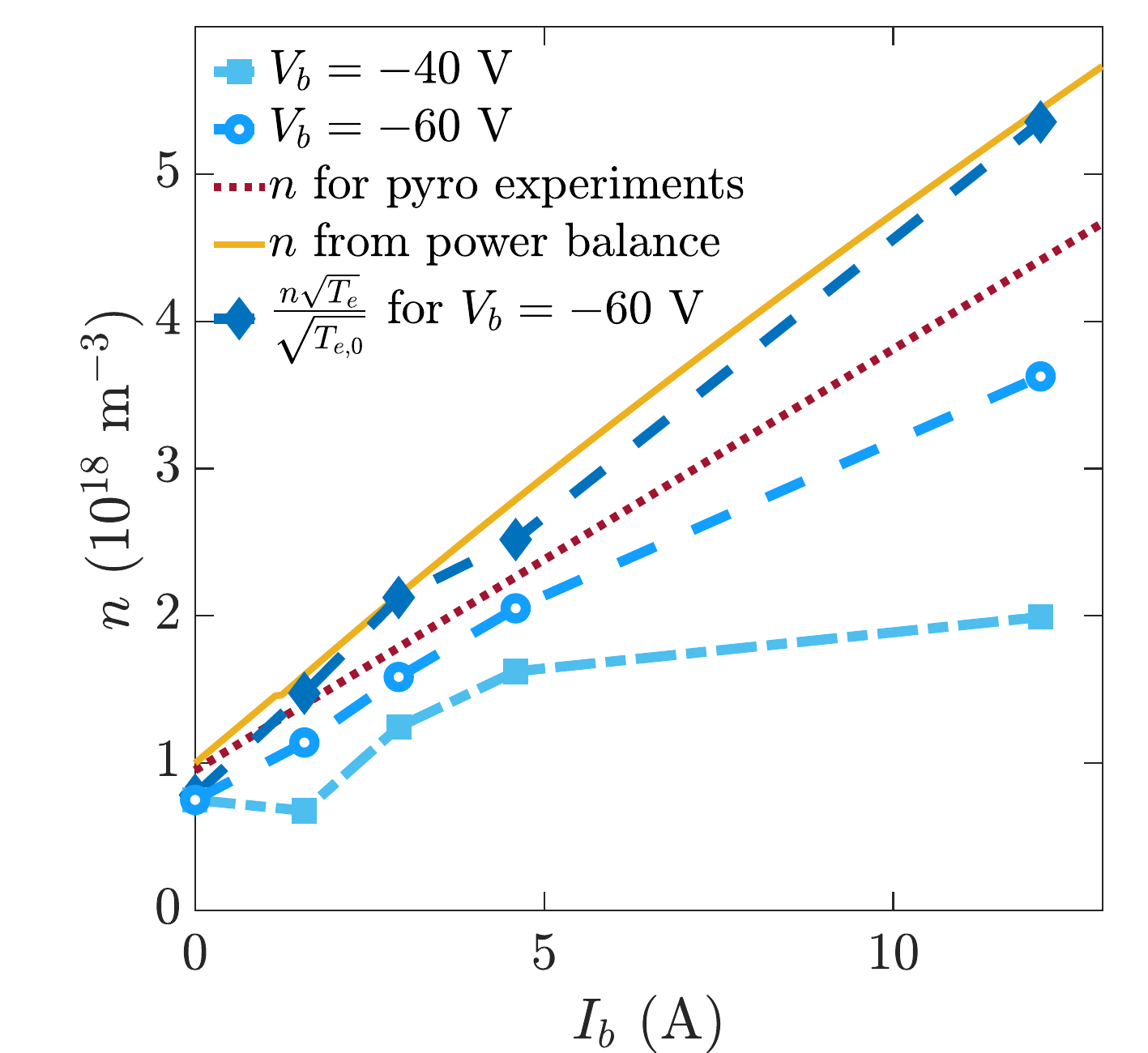}
    \caption{Density at the plasma core as a function of current through the cathode $I_b$ for two values of $V_b$. The empirical and numerical values used to compute $I_{is}$ at $T_{e,0} = 4.5~$eV are shown respectively in dotted and solid lines. The density corrected by the experimental electron temperature for $V_b = ~–60~$V is also displayed.}
    \label{fig:n_ib}
\end{figure}

The power supplied by the cathode to the plasma contributes to ionization and heating in addition to the RF antenna, resulting in a broad increase in density at the plasma core, and a localized increase of the electron temperature at the core.
The variations of the plasma density $n$ as a function of $I_b$ are shown in \fref{fig:n_ib} for two different biases of the cathode ): $V_b = ~–60~$V (circles) for $V_b = ~–40~$V (squares). 
The influence of the increase of electron temperature on the ion saturation current is also assessed in \fref{fig:n_ib} for $V_b = ~–60~$V, where $n\sqrt{T_e}/\sqrt{T_{e,0}}$ is shown as full diamond, with $T_{e,0} = 4.5~$eV the temperature at $I_b=0$.

Let us first discuss the evolution of the plasma density taken into account for the estimate of the ion saturation current in Section 4, assuming that the electron temperature remains constant at $T_{e,0} = 4.5~$eV. The linear evolution considered in Section 4 is shown as a dashed red line ($n = 10^{18} + 3 \times 10^{17}I_{em}$). 
This estimated density increase slightly overestimates the density increase (nearly $20~$\%) and slightly underestimates the real ion saturation current (nearly $20~$\%). Since the contribution of the ion saturation current on the total cathode current is minor in regards with thermionic emission, we consider that our linear modeling is accurate enough for the computation of the ion saturation current with $I_b$ as computed in Section 4.

Let us now discuss further the power balance used in the model of Section 5. 
When the cathode is left floating or cold (i.e. $I_b = 0$), the plasma density $n_0$ is sustained by the RF antenna alone and is assumed to be homogeneous in a cylinder of radius $R = 5~$cm. The density increase due to the cathode is assumed to be linear from the edge of the plasma ($5~$cm) to the core  ($0~$cm) accordingly to the experimental measurements (see \fref{fig:n_r}) and the plasma density is modelled as $n(r) = n_0 + \Delta n \left( 1- \frac{r}{R}\right)$. $T_e$ is assumed to be independent of the cathode current and cathode bias voltage. The global model assumes that the total absorbed power (i.e. the radio-frequency power and the cathode power) is balanced by the power losses at the walls, \textit{i.e.} the recombination of each electron-ion pair releases $\mathcal{E}_\mathrm{T} = 75~$eV~\cite{lieberman,ChabertBraithwaite}.

The power input from the cathode to the plasma is provided by thermionic electrons as: 
\begin{equation}
    P_{cath} = \int_0^{l_W}I_{em}(x)(\Phi_p-V_c(x))~\mathrm{d}x
\end{equation}

On the other hand, the increase in losses $\delta P_{losses}$ is associated to the density increase  $\Delta n \left( 1- \frac{r}{R}\right)$ as

\begin{equation}
    \delta P_{losses} = \Delta n \frac{\pi R^2}{3}\sqrt{eT_e/m_i}e\mathcal{E}_\mathrm{T}
\end{equation}

Assuming stationarity, one can compute the density at the core ($r = 0~$cm) as: 

\begin{equation}
    n = n_0 + \frac{3P_{cath}}{\pi R^2\sqrt{eT_e/m_i}e\mathcal{E}_\mathrm{T}}
\end{equation}

\noindent which is shown as a full yellow line in \fref{fig:n_ib} and reproduces well the experimental data corrected in $T_e$.

\section{Technical details on the numerical simulation}
\label{sec:details}

The numerical simulation solves an integro-differential heat equation. The time step is $10^{-3}~$s and the spatial step is $10^{-3}~$m. This resolution verifies the Courant-Friedrich-Levy condition for the heat equation 

\begin{equation}
    \frac{\lambda_W}{\rho C_p}\frac{\mathrm{d}t}{\mathrm{d}x^2} \sim 2.5\times 10^{-2}\leq 1
\end{equation}

For the computation of iteration N, the integral term is computed manually from the temperatures at N-1. The equation is solved numerically using the function \verb|pdesolver| from MATLAB. The initial temperature profile of the cathode is obtained from an initially homogeneous profile, then freely evolving in absence of plasma for 5 seconds.

\section{Temperature dependence of tungsten thermophysical properties}
\label{sec:properties}

\begin{figure*}[htbp]
    \centering
    \includegraphics[width = 0.95\textwidth]{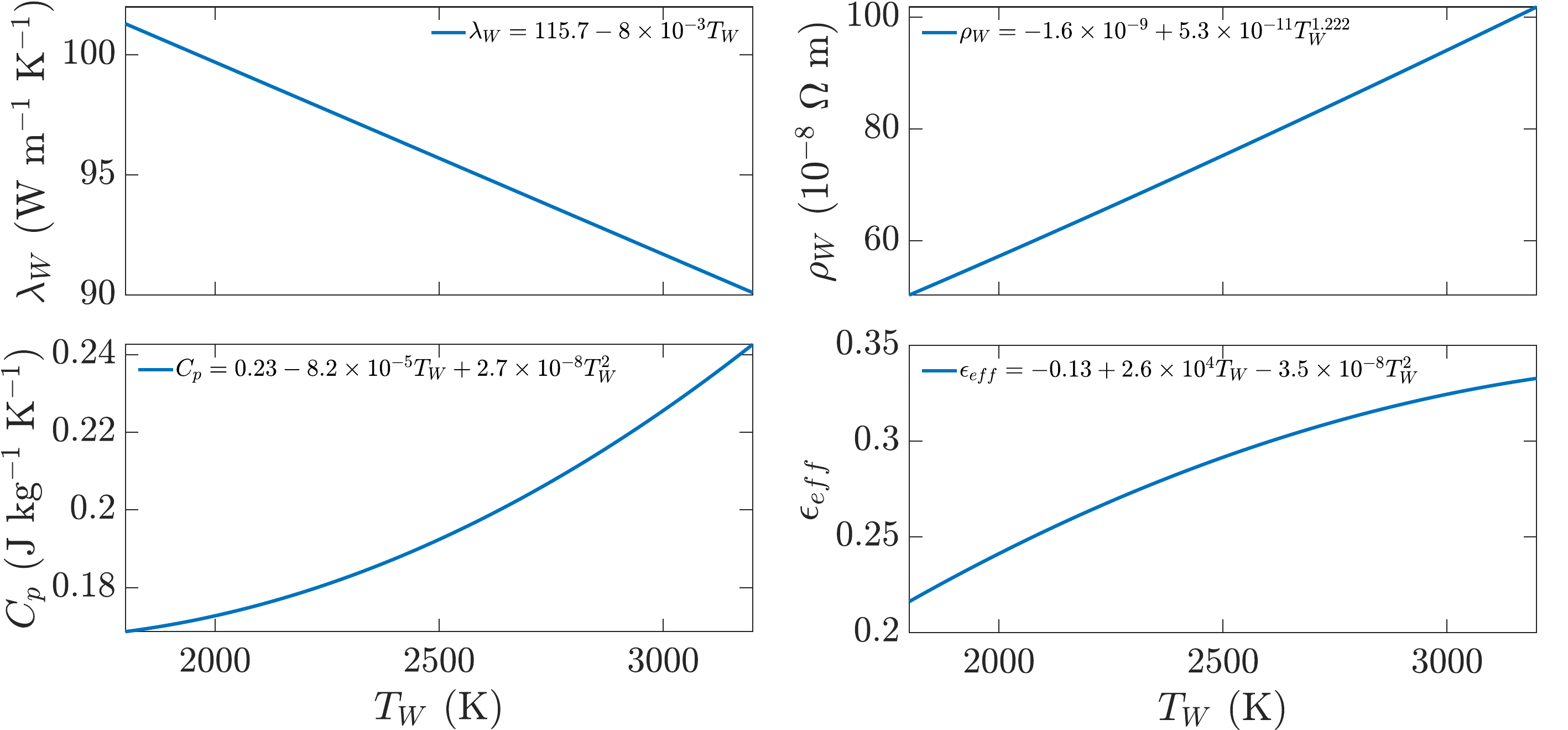}
    \caption{Thermophysical properties of tungsten as a function of temperature.}
    \label{fig:coeff}
\end{figure*}

The temperature-dependent thermophysical properties for tungsten are displayed in \fref{fig:coeff}:  thermal conductivity $\lambda_W$ \cite{white_thermophysical_1997, desai_electrical_1984}, electrical resistivity $\rho_W$ \cite{white_thermophysical_1997, desai_electrical_1984},  specific heat $C_p$ \cite{white_thermophysical_1997, desai_electrical_1984} and hemispherical total emissivity $\epsilon_{eff}$ \cite{Matsumoto1999HemisphericalTE} from top left to bottom right. The first three quantities are fits based on experimental data in the range [$1800~$K; $3200~$K] while the last one is the best fit according to the works of Matsumoto \textit{et al.} \cite{Matsumoto1999HemisphericalTE} in the range [$2000~$K; $3400~$K].

\section{Setting the value of the $\alpha$ parameter}
\label{sec:alpha}

\begin{figure}[htbp]
    \centering
    \includegraphics[width = 0.48\textwidth]{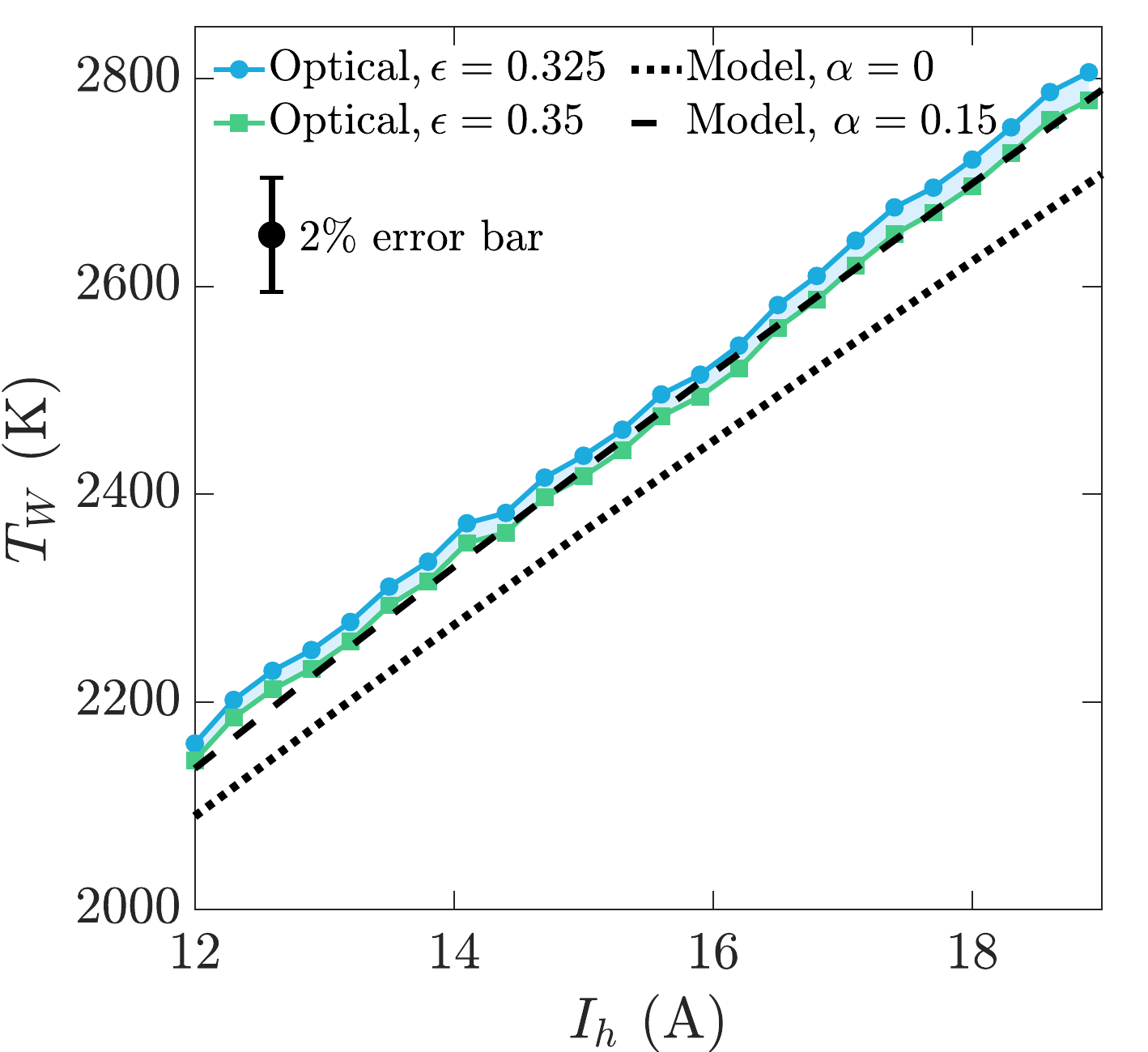}
    \caption{Comparison of optical measurements (solid lines) and numerical results (black lines) of $T_W$ for different values of $I_h$ in absence of plasma. Numerical values of $T_W$ are displayed with $\alpha = 0$ (dotted line) and $\alpha = 0.15$ (dashed line). The error bar represents the margin of error for optical measurements.}
    \label{fig:alpha}
\end{figure}

The temperature of the cathode in absence of plasma has been measured experimentally and compared to numerical results. The parameter $\alpha$ accounts for heating of the neighboring turns of the filament through radiation as explained in Figure 6. The evolution of the (homogeneous) temperature $T_W$ as a function of $I_h$ is shown in \fref{fig:alpha}. The value of the parameter $\alpha$ is  inferred from the best agreement between the experimental values and the numerical solution of the model described in Section 5. The case $\alpha = 0$ (black dotted line) underestimates $T_W$ by 50 K to 75 K  ($\sim 2.5~\%$ in relative value). A very good agreement is observed for  $\alpha = 0.15$ (black dashed line), over the whole range of $I_h$. This latter value was retained for all numerical simulations.

\section{Influence of $\lambda_{Cu}$ in the boundary conditions}
\label{sec:lambda}

\begin{figure*}[htbp]
    \centering
    \includegraphics[width = 0.95\textwidth]{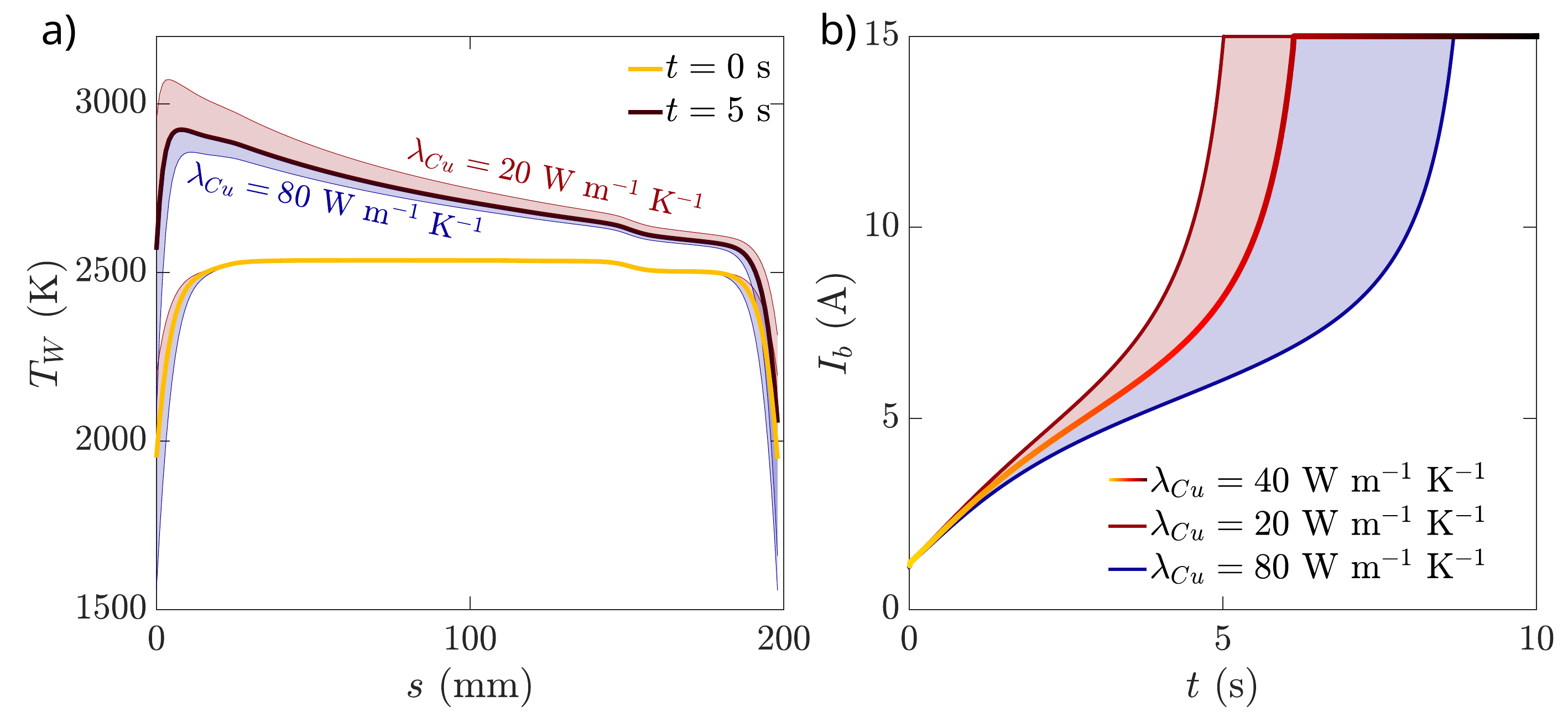}
    \caption{\textit{a)} Temperature profile along the cathode ; \textit{b)} Currents $I_b$ for 3 different values of $\lambda_{Cu}$.}
    \label{fig:lambda}
\end{figure*}

The conductivity of copper is involved in the boundary conditions as described in section 5.1.2. Its value is corrected because of the poor thermal contacts between the copper rods that are used to clamp the cathode and the tungsten filament. Its influence on the temperature profile along the cathode is shown in \fref{fig:lambda}(a) and the consequences on $I_b$ is displayed in \fref{fig:lambda}(b) for $\lambda_{Cu}$ in the range [20 W/(m K) ; 80 W/(m K)]. The blue area represents the spans of $T_W$ and $I_b$ for higher values of $\lambda_{Cu}$ up to 80 W/(m K) while the red one represents $T_W$ and $I_b$ for lower $\lambda_{Cu}$ down to 20 W/(m K). 

One can see that the initial temperature profile along the cathode is not modified except at the 15-mm extremities, where a significant difference is observed. It results in similar $I_b$ trends which differ progressively because of the small variations in initial conditions and the different losses at the boundaries. Yet this effect remains minor, as the trends in $T_W$ and $I_b$ are still similar despite spanning a wide range of $\lambda_{Cu}$.

\section{$\Phi_p$ dependency on $I_b$}
\label{sec:phi_p}

\begin{figure}[htbp]
    \centering
    \includegraphics[width = 0.48\textwidth]{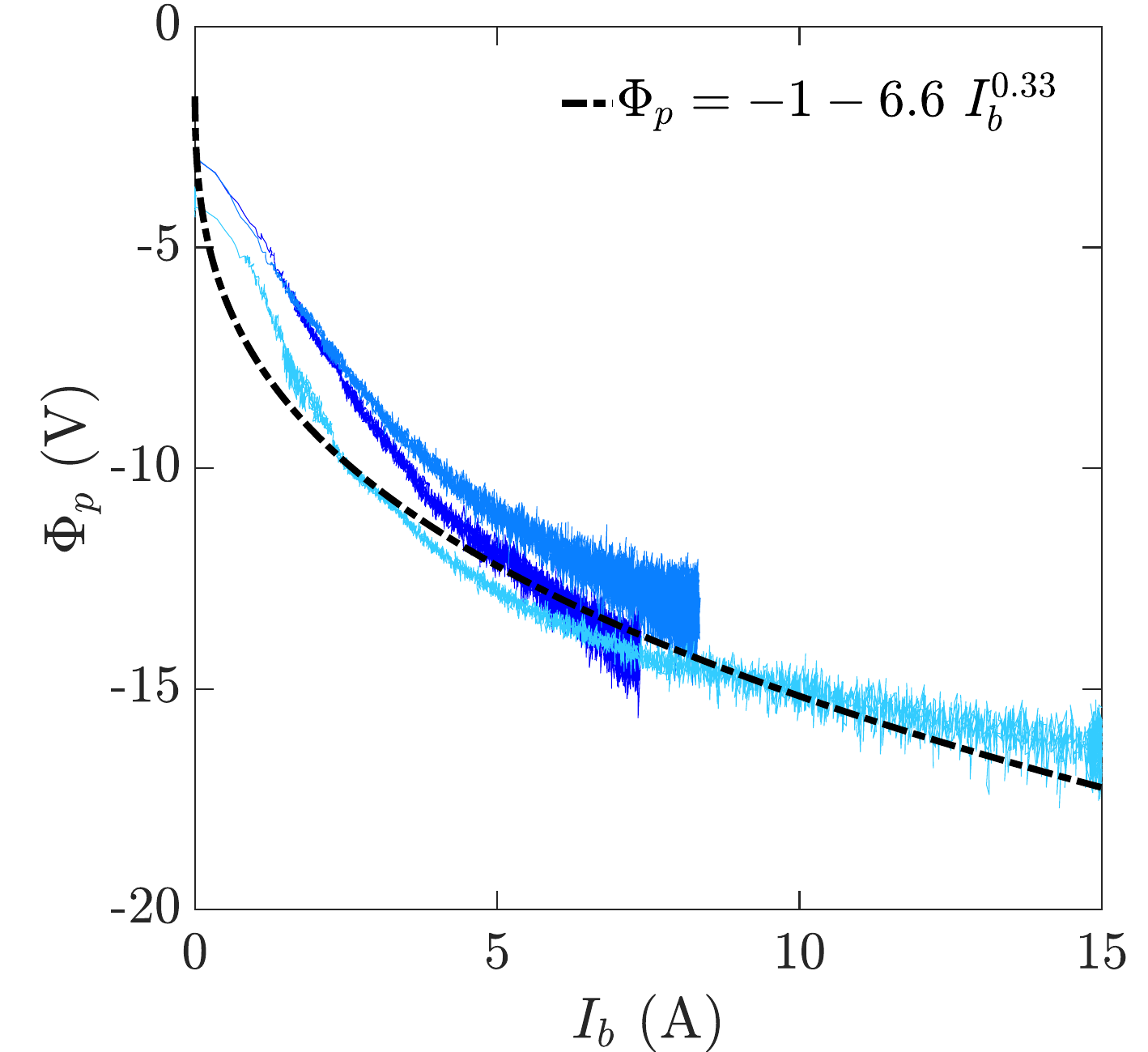}
    \caption{$\Phi_p$ measurements as a function of $I_b$. The black dotted line represents the best power law fit for the light blue curve.}
    \label{fig:vp_vs_ib}
\end{figure}

Measurements of $\Phi_p$ at the core of the plasma with an emissive probe as a function of $I_b$ are shown in \fref{fig:vp_vs_ib} for three different experiments (blue solid lines). Besides the technical challenges of using emissive probes in a non-stationary plasma, the dependence of $\Phi_p$ on $I_b$ is not fully understood yet. Its variations are thus implemented empirically in the numerical simulation. The best power law fit used for the simulations is presented here as the black dotted line.  

\section{Influence of $\beta$}
\label{sec:beta}

\begin{figure}[htbp]
    \centering
    \includegraphics[width = 0.48\textwidth]{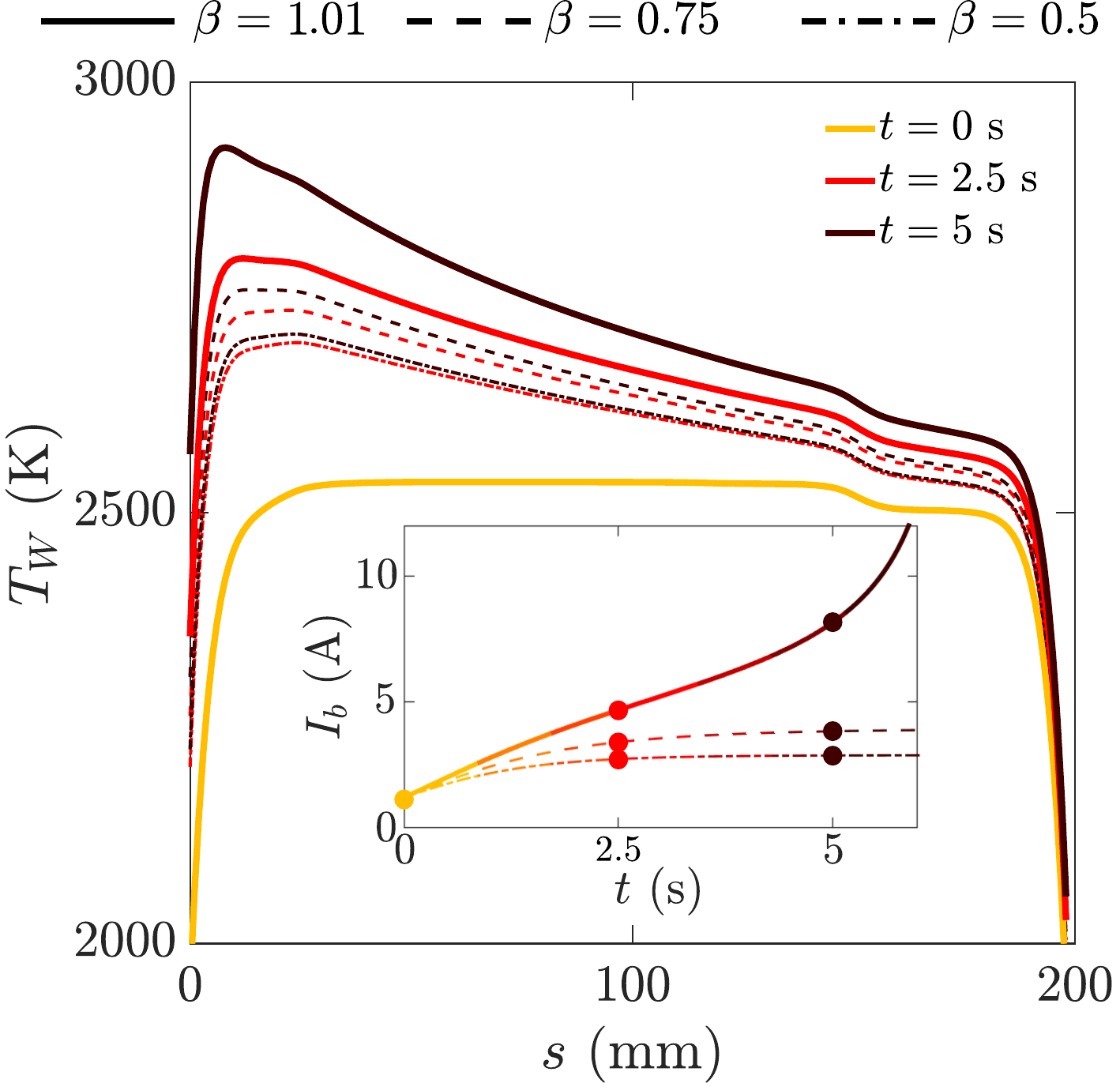}
    \caption{Temperature profiles of the cathode at $I_h = 16.2~$A and $V_b = ~–60~$V for $\beta = 0.5$ (dotted lines), $\beta = 0.75$ (dashed lines) and $\beta = 1.01$ (solid lines). Insert: $I_b$ over time for the three cases. }
    \label{fig:beta}
\end{figure}

The most sensitive numerical parameter of the full numerical model is $\beta$, introduced in section 5.1.1. Its influence is illustrated in \fref{fig:beta} as one can see temperature profiles obtained for $\beta$ values ranging from 0.5 to 0.98. The insert shows the temporal evolution of $I_b$ for three values of $\beta$.
One clearly sees the influence of $\beta$ on the temporal evolution of the cathode current. the best agreement with the experimental data is observed for $\beta = 1.01$.

\section{Source code}
\label{sec:code}

A github repository containing the source codes for the simulation is available here: \url{https://github.com/FrancisPagaud/Emissive-Cathode-Model.git}


\end{document}